\documentclass[12pt,preprint]{aastex}
\newcommand{\etal}{et~al.\ }
\newcommand{\PVdblt}{{\rm P}\kern 0.1em{\sc v}~$\lambda\lambda 1117, 1128$}
\newcommand{\CaIIdblt}{{\rm Ca}\kern 0.1em{\sc ii}~$\lambda\lambda 3934, 3969$}
\newcommand{\AlIIIdblt}{{\rm Al}\kern 0.1em{\sc iii}~$\lambda\lambda 1854, 1862$}
\newcommand{\CIVdblt}{{\rm C}\kern 0.1em{\sc iv}~$\lambda\lambda 1548, 1551$}
\newcommand{\MgIIdblt}{{\rm Mg}\kern 0.1em{\sc ii}~$\lambda\lambda 2796, 2803$}
\newcommand{\NVdblt}{{\rm N}\kern 0.1em{\sc v}~$\lambda\lambda 1239, 1243$}  
\newcommand{\SVIdblt}{{\rm S}\kern 0.1em{\sc vi}~$\lambda\lambda 933, 944$} 
\newcommand{\OVIdblt}{{\rm O}\kern 0.1em{\sc vi}~$\lambda\lambda 1032, 1038$} 
\newcommand{\SiIIdblt}{{\rm Si}\kern 0.1em{\sc ii}~$\lambda\lambda 1190, 1193$} 
\newcommand{\SiIVdblt}{{\rm Si}\kern 0.1em{\sc iv}~$\lambda\lambda 1394, 1403$} 
\newcommand{\PV}{\hbox{{\rm P}\kern 0.1em{\sc v}}}
\newcommand{\AlI}{\hbox{{\rm Al}\kern 0.1em{\sc i}}}
\newcommand{\AlII}{\hbox{{\rm Al}\kern 0.1em{\sc ii}}}
\newcommand{\AlIII}{{\hbox{\rm Al}\kern 0.1em{\sc iii}}}
\newcommand{\CaII}{\hbox{{\rm Ca}\kern 0.1em{\sc ii}}}
\newcommand{\CII}{\hbox{{\rm C}\kern 0.1em{\sc ii}}}
\newcommand{\CIIe}{\hbox{{\rm C$^{\ast}$}\kern 0.1em{\sc ii}}}
\newcommand{\CIII}{\hbox{{\rm C}\kern 0.1em{\sc iii}}}
\newcommand{\CIV}{\hbox{{\rm C}\kern 0.1em{\sc iv}}}
\newcommand{\CV}{\hbox{{\rm C}\kern 0.1em{\sc v}}}
\newcommand{\HI}{\hbox{{\rm H}\kern 0.1em{\sc i}}}
\newcommand{\HII}{\hbox{{\rm H}\kern 0.1em{\sc ii}}}
\newcommand{\Lya}{\hbox{{\rm Ly}\kern 0.1em$\alpha$}}
\newcommand{\Lyb}{\hbox{{\rm Ly}\kern 0.1em$\beta$}}
\newcommand{\Lyg}{\hbox{{\rm Ly}\kern 0.1em$\gamma$}}
\newcommand{\Lyd}{\hbox{{\rm Ly}\kern 0.1em$\delta$}}
\newcommand{\Lye}{\hbox{{\rm Ly}\kern 0.1em$\epsilon$}}
\newcommand{\Lyphi}{\hbox{{\rm Ly}\kern 0.1em$\phi$}}
\newcommand{\Lyfive}{\hbox{{\rm Ly}\kern 0.1em$5$}}
\newcommand{\Lysix}{\hbox{{\rm Ly}\kern 0.1em$6$}}
\newcommand{\Lyseven}{\hbox{{\rm Ly}\kern 0.1em$7$}}
\newcommand{\Lyeight}{\hbox{{\rm Ly}\kern 0.1em$8$}}
\newcommand{\Lynine}{\hbox{{\rm Ly}\kern 0.1em$9$}}
\newcommand{\Lyten}{\hbox{{\rm Ly}\kern 0.1em$10$}}
\newcommand{\Lyeleven}{\hbox{{\rm Ly}\kern 0.1em$11$}}
\newcommand{\HeI}{\hbox{{\rm He}\kern 0.1em{\sc i}}}
\newcommand{\HeII}{\hbox{{\rm He}\kern 0.1em{\sc ii}}}
\newcommand{\FeI}{\hbox{{\rm Fe}\kern 0.1em{\sc i}}}
\newcommand{\FeII}{\hbox{{\rm Fe}\kern 0.1em{\sc ii}}}
\newcommand{\FeIII}{\hbox{{\rm Fe}\kern 0.1em{\sc iii}}}
\newcommand{\MnII}{\hbox{{\rm Mn}\kern 0.1em{\sc ii}}}
\newcommand{\MgI}{\hbox{{\rm Mg}\kern 0.1em{\sc i}}}
\newcommand{\MgII}{\hbox{{\rm Mg}\kern 0.1em{\sc ii}}}
\newcommand{\MgIII}{\hbox{{\rm Mg}\kern 0.1em{\sc iii}}}
\newcommand{\NI}{\hbox{{\rm N}\kern 0.1em{\sc i}}}
\newcommand{\NII}{\hbox{{\rm N}\kern 0.1em{\sc ii}}}
\newcommand{\NIII}{\hbox{{\rm N}\kern 0.1em{\sc iii}}}
\newcommand{\NV}{\hbox{{\rm N}\kern 0.1em{\sc v}}}
\newcommand{\OVI}{\hbox{{\rm O}\kern 0.1em{\sc vi}}}
\newcommand{\OI}{\hbox{{\rm O}\kern 0.1em{\sc i}}}
\newcommand{\OII}{\hbox{{\rm O}\kern 0.1em{\sc ii}}}
\newcommand{\OIV}{\hbox{{\rm O}\kern 0.1em{\sc iv}}}
\newcommand{\SI}{{\rm S}\kern 0.1em{\sc i}}
\newcommand{\SIV}{{\rm S}\kern 0.1em{\sc iv}}
\newcommand{\SVI}{{\rm S}\kern 0.1em{\sc vi}}
\newcommand{\SiI}{\hbox{{\rm Si}\kern 0.1em{\sc i}}}
\newcommand{\SiII}{\hbox{{\rm Si}\kern 0.1em{\sc ii}}}
\newcommand{\SiIII}{\hbox{{\rm Si}\kern 0.1em{\sc iii}}}
\newcommand{\SiIV}{\hbox{{\rm Si}\kern 0.1em{\sc iv}}}
\newcommand{\SII}{\hbox{{\rm S}\kern 0.1em{\sc ii}}}
\newcommand{\SIII}{\hbox{{\rm S}\kern 0.1em{\sc iii}}}
\newcommand{\NaI}{\hbox{{\rm Na}\kern 0.1em{\sc i}}}
\newcommand{\TiII}{\hbox{{\rm Ti}\kern 0.1em{\sc ii}}}
\newcommand{\kms}{\hbox{km~s$^{-1}$}}
\newcommand{\cmsq}{\hbox{cm$^{-2}$}}
\newcommand{\cc}{\hbox{cm$^{-3}$}}
 
\begin{document}

%\received{date month year}
%\accepted{date month year}
%\journalid{number}{date month year}
%\articleid{number}{number}
%\slugcomment{submitted to: {\it The Astrophysical Journal}}
 
\shortauthors{DING ET~AL.}
\shorttitle{Absorption Signature of Six {\MgII}--Selected Systems}

%%%%%%%%%%%%%%%%%%%%%%%%%%%%%%%%%%%%%%%%%%%%%%%%%%%%%%%%%%%%%%%%%%%%%%%%%%%%%%%%%%
%% - Title and Author List 
%%\title{\altaffilmark{1,2}}

\title{The Absorption Signature of Six {\MgII}--Selected Systems Over $0.5 \le z \le 0.9$\altaffilmark{1,2}}

\author{Jie~Ding\altaffilmark{3}, Jane~C.~Charlton\altaffilmark{3}, and Christopher~W.~Churchill\altaffilmark{4,5}}

\altaffiltext{1}{Based in part on observations obtained at the
W.~M. Keck Observatory, which is operated as a scientific partnership
among Caltech, the University of California, and NASA. The Observatory
was made possible by the generous financial support of the W. M. Keck
Foundation.}
\altaffiltext{2}{Based in part on observations obtained with the
NASA/ESA {\it Hubble Space Telescope}, which is operated by the STScI
for the Association of Universities for Research in Astronomy, Inc.,
under NASA contract NAS5--26555.}
\altaffiltext{3}{Department of Astronomy and Astrophysics, The Pennsylvania State University, University Park, PA 16802, {\it ding, charlton@astro.psu.edu}}
\altaffiltext{4}{Department of Astronomy, New Mexico State University
  1320 Frenger Mall, Las Cruces, New Mexico 88003-8001, {\it cwc@nmsu.ed
u}}
\altaffiltext{5}{Visiting Astronomer at the W.~M. Keck Observatory}

\begin{abstract}

We present the following results of photoionization modeling of six
{\MgII}--selected absorption systems, at redshift $0.5\le z
\le0.9$, along lines of sight toward three quasars: PG~$1241+176$,
PG~$1248+401$, and PG~$1317+274$.  These are part of a larger sample
of $\sim 20$ {\MgII} absorbers at intermediate redshift, that
facilitates a survey of the properties of, and processes active in,
the multiple phases of gas, both in and outside of galaxies.  We
present new high resolution ultraviolet spectra from {\it HST}/STIS as
well as high resolution optical spectra obtained with HIRES/Keck.
Together, these spectra allow simultaneous study of a variety of
ionization states of different chemical elements, with resolution of
the kinematics providing strong constraints on multiphase properties.
The six systems presented here include several that are likely to
arise from lines of sight through disk and halo structures of spiral
galaxies.  For these we find that the high ionization gas, traced
by {\CIV} is consistent with coronal structures that resemble that
of the Milky Way, along with high velocity clouds at $100$--$200$~{\kms}.
Another system has relatively weak {\CIV}, relative to the low ionization
absorption, which leads to a discussion of the circumstances in which
a corona might be weak or absent in a galaxy, i.e. an early--type morphology
or low star formation rate.  Finally, we present systems that are examples
of single--cloud and multiple--cloud weak {\MgII} absorbers, which are
as a class not likely to be within $\sim 50$~kpc of luminous galaxies.
Based the physical conditions inferred from models, we discuss the nature
of these types of systems, which may trace outer regions of galaxies,
dwarf galaxy satellites, or metal--rich regions of the intergalactic medium.

\end{abstract}

%\pagenumbering{roman}
%\tableofcontents \newpage \newpage \pagenumbering{arabic}


\section{Introduction}
\label{sec:intro}

Absorption lines due to many different chemical transitions are
produced in quasar spectra, due to intervening absorption systems.
These systems are sampled by quasar lines of sight through galaxies and
through intergalactic gas clouds.  The absorption line spectra provide
an opportunity to sample many different types of structures within and
around galaxies.  Many of the structures that are sampled,
particularly at high redshift, emit very little, if any, light.
Nonetheless, quasar absorption lines provide a way to study detailed
properties of both visible and invisible absorbing structures.  The
kinematics, chemical composition and ionization states of absorption
systems reveal multi-phase media \citep{multiphase,1206,1634,weak1634}.

A detailed study of the properties of gaseous phases, with a range of
densities from $10^{-4}$~{\cc} to $10$~{\cc} samples environments from
the diffuse halo gas bridging galaxies to intergalactic medium to the
dense regions in galaxies where stars are forming.  It is valuable to
place a quasar absorption line of sight in the context of the
morphology of the galaxy through which it passes.  However, the
essence of quasar absorption line studies is to focus on the
properties of the absorbing gas.  This is because many of the
types of environments we are studying through the absorption
lines will not be detected through emission, particularly
at high redshift.  Ultimately, large numbers lines of
sight through galaxies, protogalaxies, galactic environments,
and intergalactic regions from redshifts $0$ to $5$ will provide
detailed evolutionary information about physical conditions.

Historically, {\MgII} absorption systems have been classified as
``strong'' or ``weak'', divided by an rest frame equivalent width
cutoff at $W_{r}(2796)=0.3$~{\AA}.  Although this division was due
primarily to the sensitivity of the previous generation of quasar
absorption line surveys \citep{ss92}, it does apparently have some
physical significance, at least roughly.  Strong {\MgII} absorbers are
nearly always found within an impact parameter of $35h^{-1}$~kpc of a
luminous galaxy ($>0.05L^*$, where $L^*$ is the Schechter luminosity)
\citep{bb91,bergeron92,lebrun93,sdp94,s95,3c336}.
Strong {\MgII} absorbers at redshifts $z=0.5$ to $0.9$ are therefore
likely candidates for predecessors of present--day giant spiral and
elliptical galaxies.

In contrast, weak {\MgII} absorbers (those with
$W_{r}(2796)<0.3$~{\AA}) do not usually have a luminous galaxy within
a $\sim 50$~kpc impact parameter (\citet{weak2} and references
therein).  Despite this, many of these weak {\MgII} absorbers have a
relatively high metallicity \citep{weak2,weak1634}, so they can be
used to trace metal production in intergalactic space and/or in dwarf
and low--surface--brightness galaxies. Therefore, they are of great
significance in the search for elusive types of star--forming
environments.

This paper presents the derived properties of the multiple phases of
gas in six {\MgII} absorption systems at $0.5 < z < 0.9$ along the
lines of sight toward three quasars.  Three of these six systems are
strong {\MgII} absorbers and the other three are weak {\MgII} absorbers.
These six systems at part of a
larger study of $21$ {\MgII} absorbers toward seven quasars for which
high resolution spectra were available both in the ultraviolet and
optical.  This facilitates coverage of many different chemical
transitions for systems at $0.4 < z < 1.4$, the redshift range which
is the focus of our study.  Our previous papers apply the same
modeling techniques and solve for the physical properties of the
multiple phases of gas in $13$ other {\MgII} absorption line systems
\citep{1634,1206,weak1634,zonak,masiero}.

As would be expected, there are a variety of absorption properties
even among strong or weak {\MgII} absorption systems.  This indicates
a range of physical conditions along different lines of sight.  This
is in part due to differences between different morphological types of
galaxies, but it is also largely due to stochastic variations from place
to place within galaxies and to evolutionary differences between different
regions.  Because of the variety, it is illuminating to further
classify {\MgII} systems according to their absorption properties in order
to understand the common processes that may be involved.
It is appropriate to focus on absorption properties rather than the
morphology of the host galaxy for several reasons: 1) Sometimes
galaxy morphology information is unavailable, especially at high redshift;
2) Much is to be learned by studying gas in the universe that is not
directly related to luminous galaxies, particularly gas in dwarf galaxies
or protogalaxies;  3) The properties of the gas are crucial for an understanding
of processes (e.g., star formation, supershells, superbubbles) that occur
globally in many different types of galaxies and environments.

\citet{archive2} developed a classification system for {\MgII} absorbers,
based on a multi--variate analysis using equivalent widths of {\CIV},
{\MgII}, {\FeII}, and {\Lya}, and the kinematic spread of a high
resolution {\MgII} profile as defining properties.  The classes of
absorbers discussed in their paper were "classic", "{\CIV}--deficient",
"single/weak", "double", and "damped {\Lya}/{\HI}--rich.
Classic absorbers have strong {\MgII} absorption, with profiles
consistent with disk/halo kinematics (a dominant component and
``satellite'' clouds), and equally strong {\CIV} absorption.
It has been suggested that much of the {\CIV} absorption is due
to a diffuse phase that is analogous to the
Galactic corona \citep{archiveletter,archive2}.
The {\CIV}--deficient absorbers have {\CIV} absorption that is
weak compared to {\MgII} and they seem to have red galaxy hosts
\citep{archive2}, perhaps suggesting that a corona is absent
in these cases.  Single/weak absorbers have weak absorption in
all transitions, including {\MgII} with $W(2796)<0.3~${\AA} and
usually with $W(2796)<0.1$~{\AA}.  Double absorbers have {\CIV}, {\MgII}, {\FeII},
and {\Lya} with double the equivalent width of classic absorbers,
such as would be expected from lines of sight through galaxy pairs
\citep{archive2,1206}.  Finally, damped {\Lya}/{\HI}--rich absorbers
have {\MgII} and {\Lya} that is double that of the classics, but
{\CIV} that is similar to the classics.

These five classes do not connect directly to particular morphological
classes of galaxies, although there may in some cases be correlations.
In some cases they do seem to connect to processes and types of structures,
e.g. the absence of a corona may lead to a {\CIV}--deficient absorber, a pair
of galaxies to a double absorber, and a high density environment near
a star forming region to a damped {\Lya}/{\HI}--rich absorber.  The goal
of this work is to learn more about the specific processes involved in
producing the various absorption signature, in order to ultimately
draw more general conclusions about evolution.

Among the six absorbers modeled in this paper, there are $2$
single/weak absorbers, $3$ classic absorbers, and
$1$ {\CIV}--deficient absorbers.  The $z=0.8954$ system toward
PG~$1241+176$ is a single--cloud, weak {\MgII} absorber, and its
metallicity and multi--phase properties will be compared to the
three single--cloud, weak {\MgII} absorbers toward PG~$1634+706$
for which the metallicity is near solar and a separate {\CIV} phase
is clearly present.  The $z=0.5584$ system toward PG~$1241+176$ is
also classified as single--weak by the multi--variate analysis
of \citet{archive2}, but it has several blended components in
{\MgII} and is similar to the $z=1.04$ system toward PG~$1634+706$
that \citet{zonak} hypothesized to be a pair of dwarf galaxies or
a superwind.  The three classic absorbers presented in this paper
(the $z=0.5504$ system toward PG~$1241+176$ and the $z=0.7729$ and
$z=0.8545$ systems toward PG~$1248+401$) have
kinematically similar {\MgII}, a dominant component with ``satellite
absorption'' reminiscent of Milky Way high velocity clouds.
The latter two systems have galaxy candidates at impact parameters
of $14 h^{-1}$ and $23 h^{-1}$~kpc, and the former is known to be
a spiral \citep{csv96}.
In our discussion of modeling results for these three ``classic''
systems we will
focus on the distribution of {\CIV} in these absorbers, relative
to the low ionization gas, considering its relevance for understanding
coronae and high velocity clouds.
Finally, the $z=0.6600$ absorber
toward PG~$1317+274$ is an example of a {\CIV}--deficient absorber.
It does have a host galaxy candidate, but at a surprisingly large
impact parameter of $72 h^{-1}$~kpc \citep{steidel02}.
The discussion in this papers will focus on the issues relevant
to these systems, with comparison to other similar systems modeled
previously.

We briefly describe the spectra we use in \S~\ref{sec:spectra} and
host galaxy images in \S~\ref{sec:galaxy}. A summary of our modeling
technique is presented in \S~\ref{sec:model}. In \S~\ref{sec:results}
we constrain physical parameters, outline the modeling results, and
give physical interpretations for each of the six systems.

\section{Data Analysis}
\label{sec:data}

\subsection{Absorption Spectra}
\label{sec:spectra}

We have included the spectroscopic observations toward three quasars
(PG~$1241+176$, PG~$1248+401$, and PG~$1317+274$).  A combination of
optical and ultraviolet spectra were used to study the six absorption
systems detected along these lines of sight. The UV spectra were
obtained with one of the primary tilts of the E$230$M grating of the {\it
Hubble Space Telescope} ({\it HST}) / Space Telescope Imaging
Spectrograph (STIS). It has a wavelength coverage from $2270$~{\AA} to
$3100$~{\AA} and a resolution of $R=30,000$ (FWHM$\sim10$~{\kms}). The
optical spectra were obtained from High Resolution Spectrograph
(HIRES) on the Keck I telescope in January 1995, with a resolution of
$R=45,000$ (FWHM$\sim6.6$~{\kms}).  Quasars, observing
dates, wavelength coverages, and total exposure times are listed in
Table~\ref{tab:spectra} for both {\it HST}/STIS and Keck/HIRES
observations.

The six systems discussed in this paper all have the low--ionization
transitions {\MgI}, {\MgII}, and {\FeII} covered in the Keck/HIRES
spectra. For the system at $z=0.5504$ toward PG~$1241+176$,
{\CaII}~$3970$ was also covered by HIRES. The {\CIV} profiles of all
six system are covered in the {\it HST}/STIS spectra. The systems at
$z=0.8954$ toward PG~$1241+176$ and $z=0.8545$ toward PG~$1248+401$
also have {\NV} covered by STIS, and four of the systems have
{\SiIV} covered. However, only the system at $z=0.8954$ toward
PG~$1241+176$ has spectral coverage for the {\Lya} absorption.

The {\it HST}/STIS spectra were reduced using the standard pipeline
\citep{brown} and combined and continuum fit using standard IRAF tasks
\citep{cv01}. The HIRES spectra were reduced with the IRAF APEXTRACT
package for echelle data and were extracted using the optimal
extraction routine of \citet{horne} and \citet{marsh}. The wavelengths
were calibrated to vacuum using the IRAF task ECIDENTIFY, and shifted
to heliocentric velocities.

\subsection{Galaxy Properties}
\label{sec:galaxy}

Three of the six absorbers have confirmed identification of host
galaxies.  As summarized in \citet{csv96}, the $z=0.5504$ system
toward PG~$1241+176$ and the system at $z=0.7729$ toward PG~$1248+401$
have confirmed host galaxies at their redshifts. The rest--frame B and
K magnitudes were determined using a combination of broadband {\it g}
($4900$/$700$), {\it R} ($6930$/$1500$), {\it i} ($8000$/$1450$), and {\it
HST}/NICMOS images of the individual quasar field. The details of
imaging and follow--up spectroscopic observations are described in
\citet{sdp94} and \citet{sd95}.  Another system at $z=0.6600$ also has
a galaxy detected at an impact parameter of $71.6~h^{-1}$~kpc. This
galaxy counterpart, which is morphologically a spiral, was detected in
the {\it HST}/WFPC2 F$702$W image of the PG~$1317+276$ field
\citep{steidel02}. Magnitudes, impact parameters, and galaxy colors
are listed in Table~\ref{tab:galaxy}.

\section{Modeling Techniques}
\label{sec:model}

The goal of our study is to place constraints on the physical
properties of the gas that is responsible for the absorption observed
in the individual systems. For each system, a Voigt profile fit was
performed on the {\MgII}~$2796$, $2803$ doublet.  The program MINFIT,
which uses a ${\chi}^2$ formalism, finds the minimum number of
components required to fit the doublet and obtains their redshifts,
column densities, and Doppler parameters \citep{thesis}.  With the
assumption that each component is produced by an individual ``cloud'',
the photoionization code Cloudy, version 94.0 \citep{cloudy01}, is run
to model individual clouds. Details of the modeling method are also
described in \citet{1634} and \citet{1206}.

For each cloud, we consider a range of metallicities and ionization
parameters. Cloudy runs to constrain the low--ionization phase are
optimized on the measured {\MgII} column density. We started with a
QSO--only spectrum \citep{hm96} as the ionizing background. An
alternative input spectrum, with the inclusion of an estimated
contribution from star--forming galaxies, was also explored for each
system.  The photon escape fraction from star--forming galaxies is
$f_{esc}=0.05$ at $z \le 3$ \citep{hm01}. A solar abundance pattern
was initially assumed, with adjustment made later, if
necessary. For each set of trial metallicity and ionization parameter,
a ``model'' spectrum is produced for each transition and is then
convolved with an instrumental spread function to generate a synthetic
spectrum. This synthetic spectrum is then superimposed on the observed
spectrum for comparison.

In many cases, the {\MgII} clouds can account for only a small
fraction of the absorption seen in the high--ionization transitions,
such as {\SiIV}, {\CIV}, and {\NV}. Thus, an additional diffuse phase
is required to give rise to these transitions. A Voigt profile fit to
the {\CIVdblt} doublet was usually performed to obtain redshift,
column density, and Doppler parameter for each needed high--ionization
cloud. In cases where {\CIV} is saturated or contaminated by
absorption from other transitions, either the {\SiIVdblt} doublet or
the {\NVdblt} doublet was used instead. Based on the fitting results
optimizing on the measured transition, Cloudy \citep{cloudy01} was run
again to constrain the metallicity and ionization parameter of the
high--ionization phase. For the high--ionization phase, both
photoionization and collisional ionization are considered as the
possible ionizing mechanisms. For collisional ionization, the
temperature in the absorbing gas (higher than the equilibrium
temperature, e.g. due to shock heating) is specified in addition to
the ionization parameter in the Cloudy code.  By following a series of
procedures similar to those described for the {\MgII} phase, we obtain
synthetic spectra for the high--ionization phase.

By constraining metallicity, ionization parameter, and/or kinetic
temperature (for collisional ionization), we try to obtain the minimum
number of phases (i.e. groups of gaseous clouds with similar densities
and temperatures) that are required to reproduce the observed spectrum
for various transitions. Alternative spectral shapes and abundance
patterns are also explored, as mentioned earlier.

\section{Modeling Results and Discussion}
\label{sec:results}

Modeling results, with ranges of acceptable parameters, are presented
here for each of the six individual systems.  Figures~\ref{fig:0.8954}
through
\ref{fig:0.6600} show the examples of ``best models'' superimposed on
the observed spectra. Physical parameters of the plotted models are
listed in Tables~\ref{tab:0.8954}--\ref{tab:0.6600}.
For each system, the presentation of results is followed by a
discussion of the implications.

\subsection{The $z=0.8954$ System Toward PG~$1241+176$}
\label{sec:z0.8954}

\subsubsection{Results}
\label{sec:r0.8954}

\begin{figure*}
\figurenum{1} 
\epsscale{0.8} 
\plotone{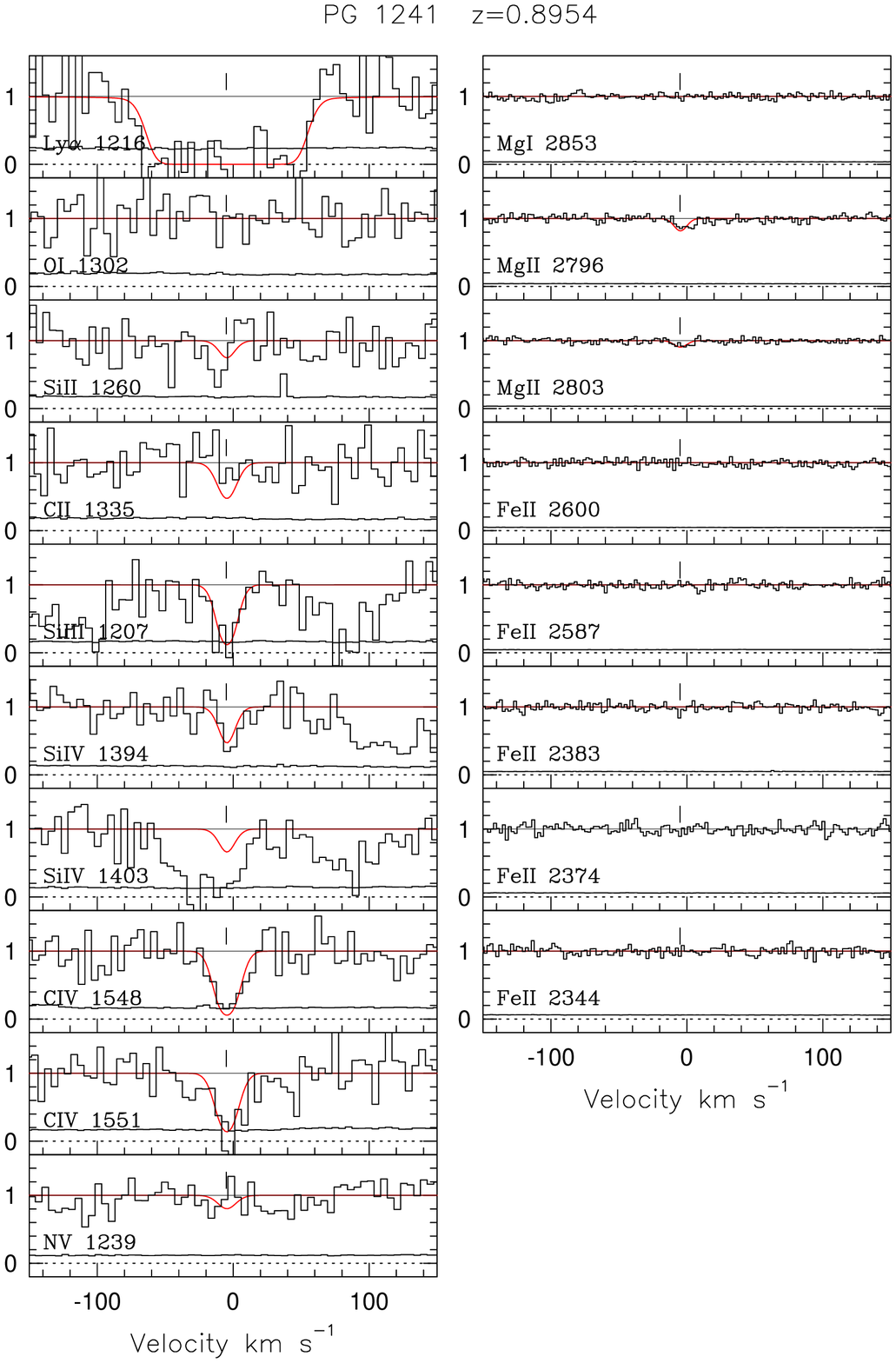}
\protect
\caption{Various key transitions are presented in velocity space, with
the zero velocity set at $z=0.895484$.The {\MgI}, {\MgII}, and {\FeII}
profiles were observed with Keck/HIRES ($R=45,000$), and all the
others with {\it HST}/STIS ($R=30,000$). The solid histogram just
above zero represents the $1\sigma$ error spectrum. The ticks in
the lower row mark the locations of {\MgII} components, resulting from
a Voigt profile fit to the {\MgII}. The ticks in the upper row
represent components in the high--ionization transitions. The solid
curve superimposed on the spectrum is the synthesized model fit, as
summarized in Table~\ref{tab:0.8954}.
\baselineskip = 0.7\baselineskip
\scriptsize{} }
\label{fig:0.8954}
\end{figure*}

\begin{figure*}
\figurenum{2}
\epsscale{0.8}
\plotone{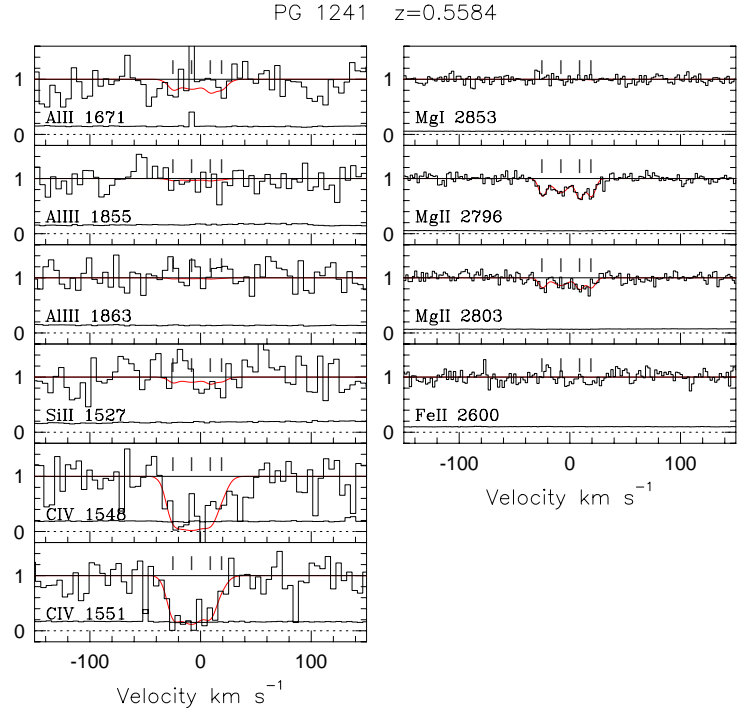}  
\protect\caption{
%\baselineskip = 0.7\baselineskip
The same as Fig~\ref{fig:0.8954}, except for the system at $z=0.5584$
toward PG~$1241+176$.}
\label{fig:0.5584}
\end{figure*}

\begin{figure*}
\figurenum{4}
\epsscale{0.8}
\plotone{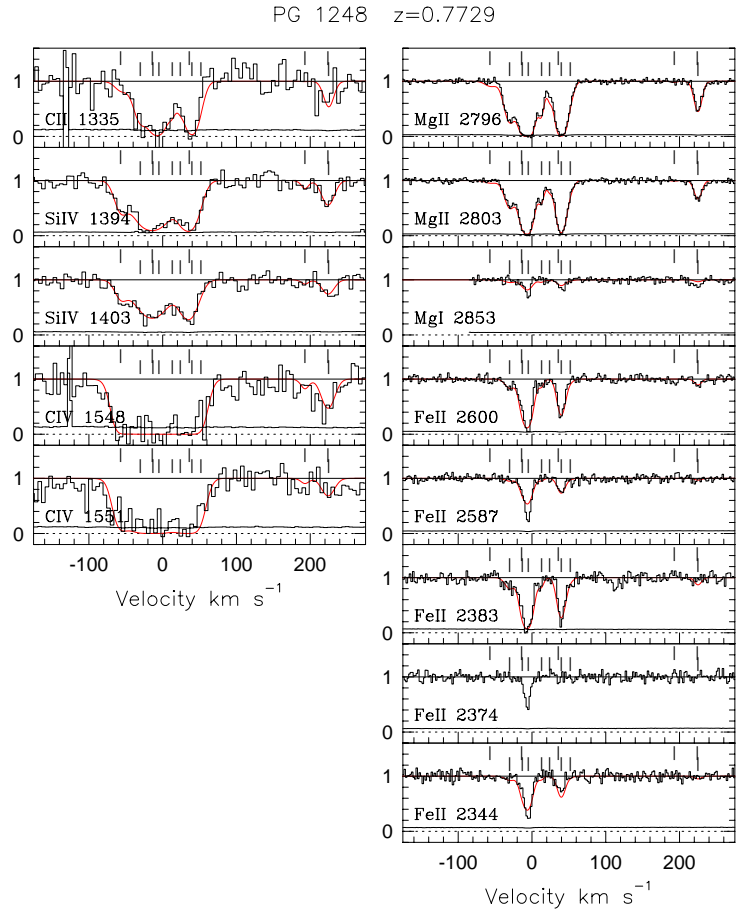}  
\protect\caption{
%\baselineskip = 0.7\baselineskip
The same as Fig~\ref{fig:0.8954}, except for the system at $z=0.7729$
Toward PG~$1248+401$.  }
\label{fig:0.7729}
\end{figure*}

\begin{figure*}
\figurenum{6}
\epsscale{0.8}
\plotone{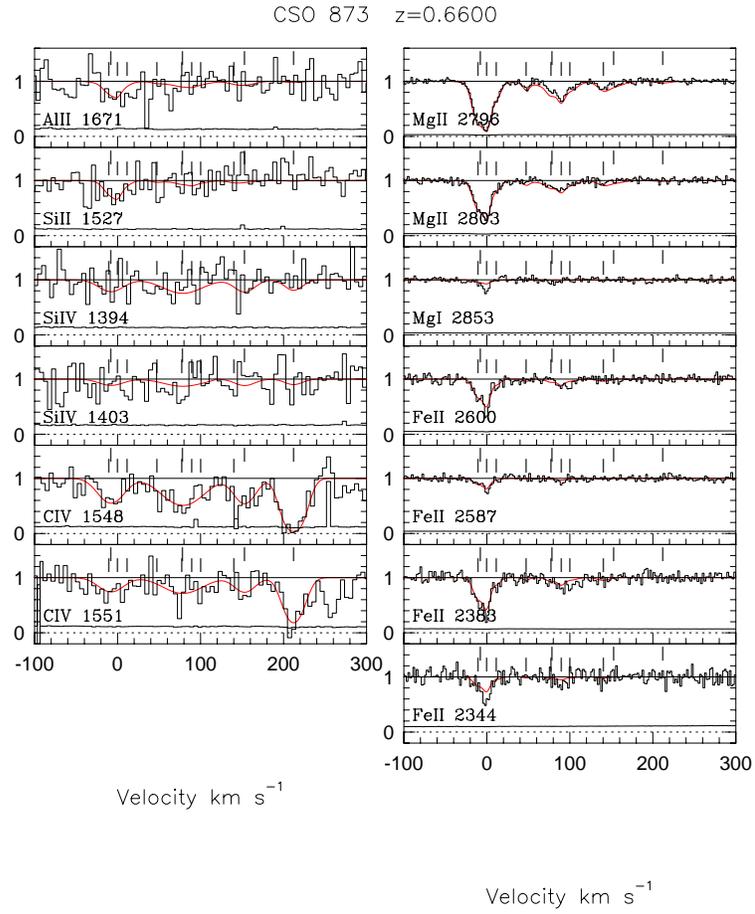}  
\protect\caption{
%\baselineskip = 0.7\baselineskip
The same as Fig~\ref{fig:0.8954}, except for the system at $z=0.6600$
toward PG~$1317+274$.}
\label{fig:0.6600}
\end{figure*}

As shown in Figure~\ref{fig:0.8954}, the system at $z=0.8954$ is a
single--cloud, weak {\MgII} system. In addition to the weak {\MgII}
doublet, other metal transitions covered in the {\it HST}/STIS
spectrum, such as {\SiIII}, {\SiIV}, and {\CIV}, also do not display a
multi--component structure in their profiles. These transitions, all
with a narrow, spiky profile shape, are aligned with each other in
velocity space. However, the velocity centroid of the strongest
absorption of these transitions is slightly offset from that of the
{\MgII} cloud.

A Voigt profile fit to the {\MgIIdblt} doublet yields a single
component at z=$0.895484$, with a column density of $\log N ({\MgII})
\sim 11.7$ and Doppler parameter of $b (Mg) \sim 7$~{\kms}.  A Voigt
profile fit to the {\CIVdblt} doublet yields a single component at
z=$0.895454$, a lower redshift than that of the {\MgII} cloud,
equivalent to $\sim5$~{\kms} to the blue. This offset in velocity is
unlikely to result from wavelength calibration, because it is not
present in the systems at $z=0.5584$ and $z=0.5504$ toward the same
quasar (see Figures~\ref{fig:0.5584} and~\ref{fig:0.5504}). However,
the {\it HST}/STIS spectrum is fairly noisy, especially in the regions
{\SiII}, {\SiIII}, {\SiIV}, and {\NV}, which makes it possible that we
should disregard the small velocity offset between the {\MgIIdblt} and
{\CIVdblt} doublets. Therefore, both a one--phase scenario (which
assumes {\MgII} and {\CIV} are produced in the same gas despite the
apparent velocity offset) and a two--phase scenario (which assumes two
separate phases) are explored. Relevant parameters of both models are
listed in Table~\ref{tab:0.8954}.

In the one--phase scenario, the cloud redshift is set to z=$0.895454$,
where the strongest absorption in {\SiIII}, {\SiIV}, and {\CIV}
occurs. This is also consistent with the minimum of the broad {\Lya}
profile, which is centered on the other transitions covered in the
{\it HST}/STIS spectrum, but not with the strongest {\MgII}
absorption. A profile fit to the {\CIVdblt} doublet yields a column
density of $\log N ({\CIV}) \sim 14.1$ and Doppler parameter of $b (C)
\sim 7$~{\kms}. A solar abundance pattern and a QSO--only input
spectrum are initially assumed, with the alternative input spectrum
explored later.  The ionization parameter of this cloud is stringently
constrained to be $\log U \simeq -2.1$ (within a $0.1$~dex of
uncertainty), by the observed ratio of $N({\MgII})$/$N({\CIV})$
(assuming they are produced in the same cloud despite the apparent
velocity offset). A model with this ionization parameter also gives
rise to the absorption consistent with the observed {\SiIII} and
{\SiIV}~$\lambda~1394$. It does not overproduce {\NV}, which is not
detected in the spectrum. In Figure~\ref{fig:0.8954}, it is shown that
{\SiIV}~$\lambda~1403$ is blended (possibly with {\Lyb} from a system
at $z\sim1.6$) and the spectrum quality is low for
{\SiII}~$\lambda~1260$. Therefore, the model fit to these two
transitions is not considered. At this ionization parameter, the
metallicity is equally well constrained to be $\log Z \simeq -1.7$
(within $0.1$~dex of uncertainty), by the shape of the {\Lya} profile
(assuming that only this one cloud is responsible for the full {\Lya}
absorption).  The corresponding cloud size is $\sim100$~kpc.

An alternative is a two--phase model in which the gas that produces
the majority of the {\MgII} absorption is separate from the gas that
gives rise to the other metal transitions. The results of the Voigt
profile fits to {\MgII} and to {\CIV} (separately) are used to set
redshifts and optimizing column densities for the two clouds. The
cloud centered on {\MgII} (the {\MgII} cloud hereafter) is required to
have an upper limit of $\log U \le -3$ so that it does not give rise
to {\CIV}.  For the cloud centered on {\CIV} (the {\CIV} cloud
hereafter), a lower limit of $\log U \ge -2$ is placed on the
ionization parameter, under the assumption that this cloud does not
produce a significant amount of {\MgII} at this offset velocity. An
upper limit of $\log U \le -1.7$ applies so that {\NV} is not
overproduced. The {\CIV} cloud also dominates the production of {\Lya}
and its metallicity is determined to be $\log Z \simeq -1.7$. A higher
metallicity cannot fit the blue wing of {\Lya} while a significantly
lower metallicity would cause the cloud to be Jeans unstable. The
ionization parameter and metallicity of the {\CIV} cloud in this
two--phase model do not differ significantly from what is derived for
the one--phase model. However, if the {\MgII} and {\CIV} clouds in
this two--phase model are not assumed to have the same metallicity,
then it is plausible for the {\MgII} cloud to have a much higher
(e.g. even super--solar) metallicity.

The alternative input spectrum, with the inclusion of stellar
contribution from galaxies in addition to the radiation from
background quasars \citep{hm01}, is explored for both models.  The
ionization parameter of the {\CIV} cloud in both cases would need to
be increased by $\sim0.3$~dex so as not to overproduce low--ionization
transitions, {\MgII} in particular. The need for an increase of
ionization parameter results from the sharp decrease in the number of
photons between $3$--$4$~Ryd in the stellar spectrum as compared to
the QSO--only contribution.  Given that the ionization energy from
{\CIII} to {\CIV} is $47$~eV, it follows that less {\CIV} would be
produced from a lower--ionization state. In other words, a larger
fraction of metals will be in a lower--ionization state. Therefore, in
order not to overproduce the low--ionization transitions, a higher
ionization parameter ($\log U \simeq -1.8$) would be required. For the
{\MgII} cloud in the two--phase model, the change resulting from the
background radiation is negligible due to the weak absorption.

\subsubsection{Discussion}
\label{sec:d0.8954}

Weak {\MgII} absorbers ({\MgII} systems with
$W_{r}(2796) \le0.3$~{\AA}) comprise at least two thirds of all
{\MgII} absorbers. Among the weak {\MgII} absorber population, about
two thirds have only a single, narrow {\MgII} cloud. The
single--cloud, weak {\MgII} systems usually have a more highly
ionized, diffuse component producing absorption at the same velocity
of the low--ionization {\MgII} cloud.

The system at $z=0.8954$ is a single--cloud, weak {\MgII} absorber. It
has such weak {\MgII} absorption ($W_{r}(2796)=0.018$~{\AA}) that it
is among the weakest detected \citep{weak1}. This system is also
unusual in that the {\MgII} absorption appears to be slightly offset
in velocity from the other detected transitions, such as
{\SiIV} and {\CIV}.  Both one--phase and two--phase models could be
compatible depending on whether this apparent offset is considered
significant.

A study of 15 single--cloud, weak {\MgII} absorbers at $z\sim1$ in
\citet{weak2} found that at least half of them contain two or more
ionization phases of gas (and in the others the constraint was not
available). In these multi--phase absorbers, the high--ionization
phase which produces the {\CIV} absorption also contributes
substantially to the {\Lya} equivalent width. This is consistent with
what we found for our two--phase scenario, in which the
low--ionization gas only accounts for {\MgII}, but not significantly
for any other transitions.  In addition, both the results from
\citet{weak2}, based on low--resolution {\it HST}/FOS spectra
($R=1,300$), and a more recent study by \citet{weak1634} of the three
single--cloud, weak {\MgII} absorbers toward PG~$1634+706$ using
high--resolution {\it HST}/STIS spectra, found that many
single--cloud, weak {\MgII} absorbers have near--solar or even
super--solar metallicity.  These systems apparently probe
metal--enriched pockets and hence potentially trace enriched regions
in the intergalactic medium or in dwarf galaxies. For our two--phase
model, though the high--ionization {\CIV} cloud is required to have a
relatively low metallicity ($\sim2\%$ solar), the metallicity of the
{\MgII} cloud is largely unconstrained. It is possible that
this system is of quite low metallicity compared to most
single--cloud, weak {\MgII} absorbers.  However, it is also possible
that the {\MgII} cloud in the two--phase model is not associated with
the high--ionization {\CIV} gas, but instead produced in a more
metal--rich region.

An alternative scenario for the $z=0.8954$ system is a one--phase
model. This single phase gives rise to the more highly ionized
transitions such as {\SiIV} and {\CIV}, as well as the observed,
extremely weak {\MgII} absorption. Both the ionization parameter
($\log U \sim -2$) and the size ($\sim100$~kpc) of this phase resemble
those derived for the high--ionization phase in the two--phase model.
This indicates that we might actually be seeing the {\MgII}
contributed by a diffuse, high--ionization phase only, while any
denser {\MgII} phase is not penetrated along this line of sight.

A survey of the warm--hot intergalactic gas in {\OVI} at $z\sim2.5$
sampled a group of {\Lya} absorbers with $\log N({\HI})\ge15.2$ at
higher redshifts. In these sub--Lyman limit systems (LLSs), a
photoionized phase with the temperature range $20,000\le T \le40,000$
was inferred to give rise to {\Lya}, {\SiIV}, {\CIV}, and other
lower--ionization transitions, in addition to the collisional ionized
{\OVI} gas \citep{simcoe02}. The physical properties of this
photoionized phase are similar to what we derived for the {\CIV} cloud
in either the one--phase or two--phase model for the $z=0.8954$
system. This could suggest that the $z=0.8954$ system is related to
such {\OVI} absorbers that were common at high redshift. The
low--ionization absorption that is detected in our one-- or two--phase
models might be suppressed at higher redshift due to the increased
ionizing photon density.

On the other hand, in the local universe, a study of the relationship
between a {\Lya} absorber (at $z=0.0053$) and a nearby dwarf
post--starburst galaxy in the $3$C~$273$ sightline was made by
\citet{stocke04}. The {\Lya} absorber was found similar to the three weak
{\MgII} systems toward PG~$1634+706$ modeled in \citet{weak1634},
with an {\HI} column density of $\log N({\HI})\sim15.9$, but with a
somewhat lower metallicity of $\sim6\%$ solar \citep{tripp3c273}. It
is argued that the observed absorption system is produced in the gas
that is driven out by the supernovae explosion in the nearby dwarf
galaxy
\citep{stocke04}. The {\MgII} absorption in the $z=0.0053$ system toward 
$3$C~$273$ is extremely weak ($\le45$~m{\AA} as measured by GHRS) even
when compared to that in the $z=0.8954$ system.  But with the decrease
of the background radiation from $z=1$ to $z=0$ taken into account, a
system with stronger {\MgII} and weaker {\CIV} absorption would be
expected at $z\sim0$. In this sense, the $z=0.0053$ system could be a
$z\sim0$ analog to the $z=0.8954$ system.  Notably, the $z=0.0053$
system also has an unusual offset between some of the different
transitions. Since at redshift $z=0.8954$, dwarf galaxies are too
faint to be observed, we cannot rule out the possibility that they are
responsible for the absorption seen in some sub--LLSs.  In fact, the
low metallicity of the {\CIV} phase in the $z=0.8954$ system
($\sim2\%$ solar) is consistent with the interpretation that it
originated in the ejecta from a dwarf galaxy. That is, either the
dwarf galaxy itself creates the wind with overall low metallicity or
the wind has swept up a large amount of the metal--poor gas
\citep{stocke04}.

Our results for $z=0.8954$ system toward PG~$1241+176$ lead to the
question of whether there are two different types of single--cloud,
weak {\MgII} absorbers.  The first class would have close to solar metallicity
in both the low-- and high--ionization phases, e.g. like the $z=0.8181$
and $z=0.9056$ absorbers toward PG~$1634+706$ \citep{weak1634},
and could perhaps arise through ``in--situ'' star formation in
intergalactic regions.  The second class would have lower metallicity
and could be related to dwarf galaxies or dwarf galaxy winds, e.g.
like the $z=0.0053$ absorber toward $3$C~$273$.  It remains uncertain
in which of these possible classes the $z=0.8954$ absorber belongs.

\subsection{The $z=0.5584$ System Toward PG~$1241+176$}
\label{sec:z0.5584}

\subsubsection{Results}
\label{sec:r0.5584}

As shown in Figure~\ref{fig:0.5584}, the system at z=0.5584 is a
multi--cloud, weak {\MgII} absorber. The blended {\MgII} absorption
spreads between v~$\sim-20$ and $20$~{\kms}. However, neither {\MgI}
nor {\FeII} has detected absorption to a $3\sigma$ rest--frame
equivalent width limit of $\sim0.1$~{\AA} \citep{weak1} over this
velocity range. The STIS spectrum covers {\AlII}, {\AlIII},
{\SiII}~$1527$, and {\CIV}, with only {\CIV} having detected
absorption. The {\CIV} absorption covers a similar velocity range as
{\MgII}. Therefore, in principle, this system could be a single--phase
absorber with {\MgII} and {\CIV} produced in the same phase of gas.

A Voigt profile fit to the {\MgII} doublet yielded four clouds at
v$\sim-25$, $-8$, $9$, and $19$~{\kms}. Table~\ref{tab:0.5584} lists
their redshifts, column densities, and Doppler parameters. Initially,
the radiation from background quasars is assumed to be the only
ionizing source. With the assumption that the clouds are optically
thin and all have the same ionization parameter, the absence of
detected {\FeII} places a lower limit of $\log U \ge -4$. If {\CIV}
arises in the same phase as the {\MgII} (the one--phase scenario),
then the ionization parameter can be stringently constrained by the
ratio of $N({\CIV})$/$N({\MgII})$ for each cloud. For this case, the
clouds at v$\sim-25$ and $-8$~{\kms} would have an ionization
parameter of $\log U \simeq -2.3$, with a $0.1$~dex uncertainty. The
clouds at v$\sim9$ and $19$~{\kms} are constrained to have $-2.7 \le
\log U \le -2.3$ and $-3 \le \log U \le -2.7$, respectively. If,
instead, {\MgII} and {\CIV} arise in two different phases (the
two--phase model), then only a lower limit can be placed on the
ionization parameter of the {\MgII} clouds, by the absence of {\FeII}
detection. Here, we favor the one--phase scenario, guided by our
general principle of minimizing the number of phases.

The ionization parameter ranges derived above are based on the
assumption that the system is optically thin. This places a lower
limit of $\log Z \ge -2$ on the system's metallicity. If, on the other
hand, the system is optically thick, then the ratio
$N({\CIV})$/$N({\MgII})$ would depend on both metallicity and
ionization parameter. Due to lack of the spectral coverage of any of
the Lyman series lines, the metallicity cannot be constrained for the
system. For simplicity, we focus on a scenario in which the system is
optically thin. Cloud sizes range from $1$--$23$~kpc in a one--phase
$\log Z = -1$ model, though they could be smaller for larger
metallicities.

An alternative input spectrum, with the inclusion of radiation from
star--forming galaxies, was also explored. For the one--phase
scenario, $N({\CIV})$/$N({\MgII})$ is slightly lower for the same
ionization parameters, due to the softer stellar spectrum between
$3$--$4$~Ryd.  This results in an increase in ionization parameter by
$\sim0.2$~dex.

\subsubsection{Discussion}
\label{sec:d0.5584}

The kinematics of multiple--cloud, weak {\MgII} systems can be cast
into two distinct sub--categories. The systems in the first category,
``kinematically spread'', have one or more dominant {\MgII} cloud(s)
and several weaker ones spread over a wider range (for example, the
$z=0.8545$ system toward PG~$1248+401$ which we will discuss in
\S~\ref{sec:z0.8545}).  The systems in the second category, known as
``kinematically compact'', are characterized by multiple weak clouds
(with comparable equivalent widths) blended together and spread over
less than $\sim100$~{\kms} in velocity space (for example, the
$z=1.0414$ system toward PG~$1634+706$ in \citet{zonak}). The
difference in kinematics between the two types of multiple--cloud,
weak {\MgII} absorbers suggests two distinct origins of the absorbing
gas. Specifically, the ``kinematically spread'' systems are found to
resemble classic, strong {\MgII} absorbers in their physical
properties and thought to arise in giant, luminous galaxies. Perhaps
the weakness of the {\MgII} absorption is a result of the line of
sight passing through a relatively sparse region of a galaxy, at a
large impact parameter, or of a relatively gas--free galaxy. On the
other hand, it has been argued that the ``kinematically compact''
systems could be associated with dwarf galaxies instead.

The system at $z=0.5584$ toward PG~$1241+706$ is a typical example of
the ``kinematically compact'', multiple--cloud, weak {\MgII}
absorbers. We find that a one--phase model is consistent with the
absorption seen from this system, in which both {\MgII} and {\CIV} are
produced in the same gas. As discussed earlier, this interpretation is
the most straightforward derivation from the observed absorption
profiles. Due to the extremely limited number of the transitions
covered in the spectrum, there could be additional phases that we are
unable to diagnose.

Another example of such ``kinematically compact'', multiple--cloud,
weak absorbers is the $z=1.0414$ system toward PG~$1634+706$, studied
by \citet{zonak}. This system, which consist of two subsystems
($\sim150$~{\kms} separated from each other in velocity space), have a
spectral coverage of {\Lya} and various metal lines, ranging from
{\MgII} to {\CIV}, {\NV}, and {\OVI}. The profiles of the {\OVI}
doublets in both subsystems are $\sim50$~{\kms} offset from the
lower--ionization transitions. In addition, this offset, broad
high--ionization phase is found to have a higher metallicity than the
metal--poor {\MgII} clouds.  It is argued that this system may either
be produced in a pair of dwarf galaxies whose halo components are
offset in velocity space from their central regions, or in two
opposite sides of a superwind around a dwarf that is undergoing a
starburst phase \citep{zonak}.  Given the similarity in kinematics as
well as in ionization state of the {\MgII} phase between this system
and our system at $z=0.5584$, it seems that the two systems could have
similar phase structure and could arise in similar environments. In
particular, the $z=0.5584$ system could be produced in the line of
sight that passes through a dwarf galaxy since a nearby luminous
counterpart has not yet been found \citep{csv96}. In a superwind
model, this system would be an example of passing through
high--density concentrations on only one side of the wind, while the
$z=1.0414$ subsystems imply passage through two sides. 

We also cannot rule out the possibility that a more highly ionized
phase, potentially traced by {\OVI}, is present in the $z=0.5584$
system.  Similar to the $z=1.0414$ system, an offset, broad {\OVI}
phase is also present in the $z=1.3430$ and $z=1.3230$ systems toward
PG~$0117+213$, both of which are ``kinematically spread''
multiple--cloud, weak {\MgII} absorbers.  Furthermore, the diffuse
{\OVI} phase in the $z=1.3430$ system requires a higher metallicity
than the {\MgII} phase \citep{masiero}.  If such an offset, more
metal--rich phase was to exist also in the $z=0.5584$ system, then it
would be a statistically significant increase in our sample of
multiple--cloud, weak {\MgII} absorbers. This presence of this phase
could potentially be one major clue to the difference in the physical
properties between multiple--cloud, weak and classic strong {\MgII}
systems. In this sense, it would be of particular significance if the
spectral coverage for {\OVI} was available, even at low resolution.

In the absence of detected {\FeII} in the spectrum of the $z=0.5584$
absorber, it is plausible that the {\MgII} and {\CIV} absorption in
$z=0.5584$ system is produced in the same phase of gas. In fact, this
is also the case for the $z=1.0414$ system toward PG~$1634+706$, in
which {\CIV} is produced in the {\MgII} clouds \citep{zonak}.  Despite
this possibility, it also would not be surprising if {\MgII} and
{\CIV} actually arise in two separate phases of gas. For example,
{\CIV} absorption could instead either originate in an {\OVI} phase
(as in the $z=1.3250$ system toward PG~$0117+213$ \citep{masiero}) or
in an additional intermediate phase (for example as in the
collisionally ionized phase in $z=1.3430$ system toward PG~$0117+213$
\citep{masiero}). Lacking the spectral coverage of {\SiIII}, {\SiIV},
and {\NV}, which usually place additional limits on kinematics,
ionization state, and ionizing mechanism of individual {\CIV} clouds,
it is hard to tell whether the {\CIV} phase is composed of multiple
narrow structures or a single broad component.  However, if the {\OVI}
phase is offset as the one in the $z=1.0414$ system, then {\CIV}
cannot be produced in the same phase as the {\OVI} in this
case. Again, we would be in a better position to constrain the {\CIV}
phase if the spectral coverage for {\OVI} was available.

\subsection{The $z=0.5504$ System Toward PG~$1241+176$}
\label{sec:z0.5504}

\subsubsection{Results}
\label{sec:r0.5504}

\begin{figure*}
\figurenum{3}
\epsscale{0.8}
\plotone{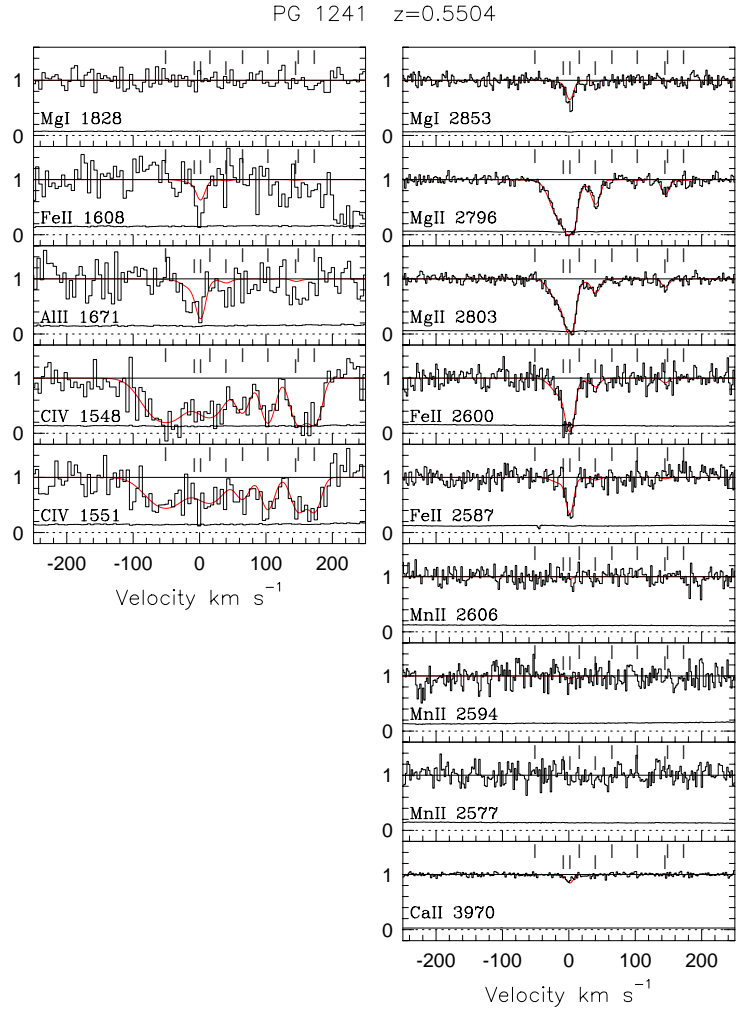}  
\protect\caption{
%\baselineskip = 0.7\baselineskip
The same as Fig~\ref{fig:0.8954}, except for the system at $z=0.5504$
toward PG~$1241+176$.  Also, {\CaII}~$3970$ is covered in the
Keck/HIRES spectrum.}
\label{fig:0.5504}
\end{figure*}

As shown in Figure~\ref{fig:0.5504}, the system at $z=0.5504$ is a
strong {\MgII} absorber, with {\MgI}, {\FeII}, and {\CaII}~3970
detected in the Keck spectrum as well. The Keck spectrum does not
cover {\CaII}~3935 at its velocity centroid where the absorption
occurs (it only covers the spectrum at v$~\ge100$~{\kms}). Aligned
with the strongest {\MgII} absorption in velocity space, is relatively
strong {\MgI} and {\FeII} absorption, both transitions having profile
shapes similar to that of the {\MgII} doublet. {\CaII}~$3970$ also has
detected absorption at $v\sim0$~{\kms}, though it is much weaker in
strength. The {\CIVdblt} doublet displays broad absorption features
that are distinctly different from those seen in the low--ionization
transitions. Hence, a multi--phase model is certainly required for
this system.

A Voigt profile fit to the {\MgII} doublet yielded four separate
clouds at v$\sim-8$, $2$, $40$, and $145$~{\kms}. Their redshifts,
column densities, and Doppler parameters are listed in
Table~\ref{tab:0.5504}. The two stronger clouds at v$\sim-8$~{\kms}
(with $\log N ({\MgII}) \sim 13.1$ and $b ({\rm Mg}) \sim 19$~{\kms})
and v$\sim2$~{\kms} (with $\log N ({\MgII}) \sim 13.5$ and $b ({\rm
Mg}) \sim 6$~{\kms}) combine to produce the saturated absorption in
{\MgII} at the system velocity zero--point.  A solar abundance pattern
and a QSO--only ionizing spectrum are initially assumed. For the two
strongest clouds at v$\sim-8$ and $2$~{\kms}, {\MgI}, {\CaII}~$3970$,
and {\FeII} were all used to constrain physical parameters. Because
the system is optically thick, the production of $N({\MgI})$,
$N({\CaII})$, and $N({\FeII})$ depend on both ionization parameter and
metallicity. For example, if the metallicity is of the solar value,
the ionization parameter is constrained by $N({\MgI})$/$N({\MgII})$ to
be $\log U\sim-5.5$; when the metallicity is one tenth of solar, the
ionization parameter is $-6.5\le \log U
\le-6$, as required by $N({\MgI})$/$N({\MgII})$. A metallicity of
$\log Z\le-2$ underproduces {\MgI} regardless of $\log U$. The
ionization parameters derived for both solar and one tenth metallicity
are both consistent with the observed $N({\FeII})$/$N({\MgII})$, but
both severely overproduce {\CaII}. It seems that an abundance pattern
adjustment with calcium decreased by $0.5$~dex relative to magnesium
is necessary for the strongest {\MgII} cloud.

An alternative approach is to create an additional phase which gives
rise to the observed {\MgI} and raise the ionization parameter of the
{\MgII} cloud such that it gives rise to less {\FeII} and {\CaII}.  A
Voigt profile fit to {\MgI}~$2853$ yields one single component, with
$N({\MgI})\sim11.85$ and $b({\rm Mg}) \sim 6.94$~{\kms}, at
v$\sim2$~{\kms}.  At one tenth solar metallicity, the ionization
parameter of this phase is constrained to be $\log U \le-8.5$, so that
{\CaII} is not significantly produced. For the dominant {\MgII} cloud
at the same velocity, the ionization parameter is constrained by
$N({\CaII})$/$N({\MgII})$ to be $\log U \ge-3.5$, so that {\CaII} is
not overproduced.  However, this ionization state range underproduces
{\FeII}. The observed $N({\FeII})$/$N({\MgII})$ requires $\log U
\le-4$. Therefore, an abundance pattern adjustment with either an
elevation of iron by $0.4$~dex or a decrease of calcium by $0.2$~dex
is needed for the strongest {\MgII} cloud.

From the two weak, more offset {\MgII} clouds at v$\sim40$ and
$145$~{\kms} (see Figure~\ref{fig:0.5504}), {\MgI}, {\CaII}, and
{\FeII} is not detected. No stringent constraint on ionization
parameter or metallicity can be obtained for the relative production
of these transitions either. Hence, for simplicity, we tabulate the
same ionization parameter and metallicity as the other stronger
{\MgII} clouds.

Regardless of whether an even lower--ionization {\MgI} phase is
included or not, only a small fraction of the observed {\CIV}
absorption could be produced in the {\MgII} phase (or {\MgI} plus
{\MgII} phases). A separate high--ionization phase is required to
produce the absorption seen in {\CIV}. A Voigt profile fit to the
{\CIV} doublet gave six distinct clouds (see Table~\ref{tab:0.5504}).
With lack of coverage of other high--ionization transitions and the
Lyman series lines, the ionization state and metallicity of this phase
are poorly constrained.  For simplicity, we assume that the six clouds
have identical physical properties (i.e. ionization state and
metallicity).  In addition, this phase is assumed to have the same
metallicity as the {\MgII} phase. Either photoionization or
collisional ionization could be the dominant ionizing mechanism for
the six clouds.  For the case of photoionization, a lower limit of
$\log U \simeq -2$ is placed on the ionization parameter such that
absorption is not significantly produced in any of the low--ionization
transitions. For the case of collisional ionization, with the
assumption of a uniform temperature, the temperature is constrained to
be $\log T \ge 3.7$ in order not to overproduce low--ionization
transitions.

A background spectrum, with the inclusion of star--forming galaxies,
is also considered for the two--phase model. The ionization parameter
of the {\MgII} phase would need to be decreased by $\sim1$~dex so that
{\MgI} would not be underproduced. The {\CIV} phase ionization
parameter does not undergo significant change, since it was poorly
constrained in the first place. An abundance decrement of
$\sim0.5$~dex in calcium is still needed for the strongest {\MgII}
cloud for a model without an additional {\MgII} phase.

\subsubsection{Discussion}
\label{sec:d0.5504}

The system at $z=0.5504$ is a ``classic'' strong {\MgII} absorber. Its low-- and
high--ionization transitions have such distinctively different
kinematics that this system is by all means a multi--phase absorber.

The $z=0.5504$ system has a nearby galaxy candidate (at an impact
parameter of $13.8~h^{-1}$~kpc). The galaxy has a B and K band
luminosities of $0.64L_{B}^{*}$ and $0.41L_{K}^{*}$, respectively, and
the rest--frame $<$B-K$>$ color of $3.42$ \citep{csv96}. The galaxy
properties are listed in Table~\ref{tab:galaxy}.

The {\MgI} absorption in the $z=0.5504$ system is particularly strong.
This provides a hint that {\MgI} could arise in a separate phase with
lower ionization than that of the {\MgII} clouds. In fact, such a
model could consistently reproduce the observed absorption profiles,
provided the ionization parameter of the {\MgII} phase is elevated as
in comparison to a one--phase model. Such an additional {\MgI} phase
is found to exist in the DLA system at $z=0.5764$ toward
PG~$0117+213$, with a density of $\sim200$--$700$~{\cc} and a
temperature of $100$'s of Kelvin \citep{masiero}, similar to our
hypothesized {\MgI} phase.  In addition, such cold, dense {\MgI}
pockets (with a density of $200$~{\cc} and a temperature on the scale
of $100$s of K) were also implied by models of a Lyman limit system
(LLS) at $z=0.9902$ toward PG~$1634+706$. These tiny concentrations of
gas are believed to be analogous to small--scale structure which is
observed in the Milky Way ISM
\citep{1634}.

Another interesting point concerning the low--ionization gas in the
$z=0.5504$ system is the relative abundance of various metal species.
As shown in the spectrum, the {\CaII} absorption is very weak in
comparison to that of {\MgI}, {\MgII}, and {\FeII}. Regardless of the
phase structure of the low--ionization gas, a decrease of the calcium
abundance by either $0.2$ or $0.5$~dex (depending on the scenario) is
needed. However, in the scenario with the inclusion of the additional
{\MgI} phase, this decrease of calcium could instead be replaced by an
elevation of iron by $0.4$~dex, with regard to magnesium and calcium.
Interestingly, a similar abundance pattern adjustment (the elevation
of iron) was reported to be necessary in both the low-- and the very
low--ionization phases in the system at $z=0.5764$ toward
PG~$0117+213$ \citep{masiero}.  This also suggests that the
low--ionization gas in the two systems is of similar origin.

A third common feature between the $z=0.5504$ system and $z=0.5764$
system toward PG~$0117+213$ is that the {\CIV} absorption cannot be
produced in the {\MgII} phase. Lacking spectral coverage of {\SiIV},
{\NV}, or {\OVI}, the physical conditions of the {\CIV} phase in
neither system can be well constrained. For the system at $z=0.5504$,
however, the narrow, resolved components apparent in the {\CIV}
profiles indicate that the high--ionization gas is unlikely to arise
in a coronal structure similar to that of the Milky Way.  The system
at $z=0.9254$ toward PG~$1206+459$ has similar narrow components
resolved in its {\CIV} and {\NV} profiles. This absorber is part of a
double {\MgII} system and is thought to arise from a patchy
distribution of gas in the outskirts of a warped spiral disk at an
impact parameter of $43~h^{-1}$~kpc \citep{1206}.

A broadband image of the PG~$1241+459$ field, in combination with
follow--up spectroscopic study, confirmed a host galaxy for the
$z=0.5504$ system at an impact parameter of $13.8~h^{-1}$~kpc
\citep{sdp94,sd95}. The galaxy has a B and K band
luminosities of $0.64L_{B}^{*}$ and $0.41L_{K}^{*}$,
respectively. With the rest--frame $<$B-K$>$ color of $3.42$, this
galaxy is likely to be a spiral (see
Table~\ref{tab:galaxy}). Therefore, it follows that the
high--ionization gas in the $z=0.5504$ system could also be produced
in the patchy corona of a mid--type spiral galaxy that does not have
too much on--going star formation activity. Resolved structures in the
high--ionization gas are not rare for strong {\MgII} absorbers. For
example, the system at $z=0.7729$ toward PG~$1248+401$ is a classic,
strong {\MgII} system.  Although its {\CIV} absorption is saturated,
there are resolved structures discerned in its {\SiIV} profiles. This
system also has a known host galaxy, with the rest--frame $<$B-K$>$
color of $3.15$ and an impact parameter of $23.2~h^{-1}$~kpc
\citep{csv96}.  Therefore, it could originate in a galaxy environment
similar to that responsible for the $z=0.5504$ absorber.

The majority of the ionized gas in the ISM of the Milky Way exists in
a vertically extended layer, known as the Reynolds layer. This warm
ionized medium (WIM) of our Galaxy, with a local midplane density
of $\sim0.1$~{\cc} and a scale height of $\sim1$~kpc, fills more than
$20\%$ of the entire ISM volume \citep{reynolds}. Such layers have
also been confirmed to exist in many other spiral galaxies by various
observing techniques, such as narrow--band emission line imaging,
long--slit spectroscopy, and Fabry--Perot observations
\citep{wb94,hwg96,fwg96}. They are more commonly referred to as
diffuse ionized gas (DIG) in external spiral galaxies.  It is believed
that radiation from OB stars within the galaxies is the primary
ionization source.  Therefore, the thickness of the DIG is dictated by
the amount of star formation in the individual galaxy. A study of four
edge--on spiral galaxies by \citet{collins01} and \citet{collins00}
indicates that the production of this layer could be consistent with
galactic fountain model, in which the dynamic halos result from
materials pushed from the midplane by the violent star formation in
the disk. It is highly possible that the patchy corona we discussed
earlier is in fact part of the DIG.

\subsection{The $z=0.7729$ System Toward PG~$1248+401$}
\label{sec:z0.7729}

\subsubsection{Results}
\label{sec:r0.7729}

As shown in Figure~\ref{fig:0.7729}, the system at $z=0.7729$ is a
multiple--cloud, strong {\MgII} system. The {\MgII} absorption spreads
over the velocity range $-50 \le v \le 70$~{\kms}, with a
high--velocity cloud located at $\sim225$~{\kms}.  The strongest
absorption seen in most low--ionization transitions, such as {\MgI},
{\FeII}, and {\CII}, is aligned with that of {\MgII} in velocity
space. The high--ionization transitions {\SiIV} and {\CIV} show strong
absorption at similar velocities, including a component centered on
the high--velocity cloud. However, the absorption in both transitions
extend further to the blue ($\sim-80$ ~{\kms}) in the spectrum. The
absorption seen in the {\CIV} doublet is strong and saturated.

A simultaneous Voigt profile fit to the {\MgII}~2796, 2803 doublet,
{\MgI}, and {\FeII} yields eight separate clouds. Their individual
column densities and Doppler parameters are listed in
Table~\ref{tab:0.7729}.  An extragalactic ionizing background due to
quasars is assumed to be the dominant ionizing source initially, while
an alternative with the inclusion of starburst galaxies will be
discussed later on.  With the assumption that the eight clouds have
the same metallicity, the ionization parameter of each cloud is
constrained by $N({\MgI})$/$N({\MgII})$ and $N({\FeII})$/$N({\MgII})$.
Specifically, the clouds have ionization parameters in the range $-5.5
\le \log U \le -4$. The only exception is the cloud at
v~$\sim24$~{\kms}, which has weak and blended absorption in {\MgI} and
{\FeII} and therefore cannot be well constrained using
$N({\MgI})$/$N({\MgII})$ and $N({\FeII})$/$N({\MgII})$. However, an
upper limit of $\log U \le -3$ is placed on the ionization parameter
of this cloud so that it does not overproduce {\CII}. Although the
ionization parameter ranges derived above are based on one tenth solar
metallicity, they are valid, within 0.5~dex, for both solar
metallicity and one hundredth solar metallicity. Lacking coverage of
the Lyman series lines, the metallicity cannot be robustly
constrained.

Regardless of the choice of metallicity, the ionization parameters of
the {\MgII} phase are so low that they produce only a small fraction
of the {\SiIV} or {\CIV} absorption at similar velocities. In
addition, both {\SiIV} and {\CIV} display strong absorption at
velocities blueward of the {\MgII} clouds. Therefore, a more highly
ionized phase is required. Because the profiles of {\CIVdblt} are
saturated, a Voigt profile fit was performed on the {\SiIVdblt}.  Five
clouds were obtained, with their individual redshifts, column
densities, and Doppler parameters listed in Table~\ref{tab:0.7729}.
These clouds are initially assumed to have the same metallicity as
that of the {\MgII} phase.

Both photoionization and collisional ionization have been explored,
with $N({\CII})$/$N({\SiIV})$ and $N({\CIV})$/$N({\SiIV})$
constraining the range for ionization parameters and kinetic
temperatures. For the case of photoionization, at $\log Z \sim -1$,
{\MgII} and {\CII} are both overproduced over the range of reasonable
ionization parameters $-2.5 \le \log U \le -1$. At $\log U \ge -1.5$,
{\CIV} is overproduced as well. Alternatively, at solar metallicity, a
reasonable fit could be obtained for most transitions when the
ionization parameter was $\log U \sim -2$, though {\CII} and {\CIV}
are both overproduced at v~$\sim193$ and ~$224$~{\kms}. An
alpha--enhancement of $\sim~0.7$~dex would be needed for these two
clouds.

For the case of collisional ionization, no temperature range that
reconciles this discrepancy could be found consistent with the data
for one tenth solar metallicity. At $\log T \le 5$, {\CII} is
overproduced while at $\log T \ge 5$, {\CIV} is overproduced.  A
change of metallicity does not allow us to find a consistent
temperature either. Given the significant overproduction of {\CII} and
{\CIV} (as compared to the photoionization case), a much greater
alpha--enhancement would be required in order for the model to
reproduce the observed profiles. Therefore, photoionization is favored
over collisional ionization for the high--ionization transitions in
this system.

We also consider the case where stellar sources make a significant
contribution to the extragalactic background radiation. The stellar
spectrum, which is softer than the extragalactic radiation at
$3$--$4$~Ryd, is less effective in ionizing {\CIII} to {\CIV} which
requires an ionizing energy of $47$~eV. Therefore, for the
high--ionization phase, this leads to the production of less {\CIV}
with similar ionization parameters. Due to the saturation of the
{\CIV} absorption between $-100 \le v \le 100$~{\kms}, this change in
the input spectrum noticeably affects only the high--velocity cloud at
v$\sim225$~{\kms}. Specifically, with a solar abundance pattern and
the same ionization parameter, {\CIV} is no longer overproduced as it
would be using the QSO--only background.  However, since {\CII} is
still slightly overproduced, the ionization parameter needs to be
increased by $\sim0.3$~dex. Still, an alpha--enhancement of $0.5$~dex
is needed due to the overproduction of {\CII} and {\CIV}. For the
low--ionization phase, the alternative input spectrum has little
influence, given that the ionizing energies of the low--ionization
transitions fall in the energy range where the contribution from O and
B stars do not make a significant difference.  Therefore, the derived
ranges for the ionization parameters are still valid for these
clouds. 

\subsubsection{Discussion}
\label{sec:d0.7729}

This system is a ``classic'' strong {\MgII} absorber with a two--phase
structure. The low--ionization clouds have a typical density of
$0.03$~{\cmsq} and sizes between $2$ and $200$~pc. The
high--ionization phase, which overlaps the low--ionization clouds in
velocity, has a typical density of $10^{-4}$~{\cmsq} and cloud sizes
between $0.2$ and $10$~kpc.  This system has a galaxy with a redshift
consistent with the absorber at an impact parameter of
$23.2~h^{-1}$~kpc from the quasar. The host galaxy has the absolute B
and K magnitudes of $0.53L_{B}^{*}$ and $0.27L_{K}^{*}$, respectively,
and the rest--frame $<$B-K$>$ color of $3.15$ \citep{csv96} (see
Table~\ref{tab:galaxy}).

The {\SiIV} and {\CIV} profiles have strong absorption extending
further to the blue in the spectrum than the majority of the
low--ionization transitions. A similar offset of the more highly
ionized transitions from the low--ionization ones is seen in another
strong {\MgII} system at z=$0.9902$ toward PG~$1634+706$ \citep{1634}.
In this system, the clouds offset to the red in the spectrum have
higher ionization parameters than the others in the low--ionization
phase and hence give rise to stronger {\SiIV} and {\CIV}, but weaker
{\MgII}.  This gradient in the ionization parameter was interpreted to
be the result of a density variation in the disk of the host galaxy
(with a range of $\sim0.06$~{\cc} to $\sim0.003$~{\cc}). However, in
the z=$0.7729$ system, the absorption in the low--ionization
transitions is so weak at the offset velocity that the ionization
parameters of the offset clouds have to be much higher than the rest
of the low--ionization clouds. This leads to the density of the offset
clouds being $\sim10^{-4}$~{\cmsq}, roughly $300$ times higher than
those of the clouds in the low--ionization phase. Therefore, it is
likely that the {\MgII} clouds arise in a medium similar to the warm,
ionized inter--cloud medium in the Milky Way \citep{3phase}, probably
the disk ISM of the luminous galaxy candidate, while the offset clouds
are produced in a separate, more diffuse phase, together with the
other clouds that are responsible for absorption seen in {\SiIV} and
{\CIV}.  This is consistent with the conclusion in
\citet{archiveletter}, where the existence and global dynamics of
smaller, kinematic "outliers" are found to be intimately linked to the
presence and physical conditions of a higher ionization phase.

Resolved structures, superimposed on the {\MgII} clouds, are seen in
the profiles of {\SiIV}. The absorption in {\CIV}, on the other hand,
is much stronger and saturated, but aligned with that of {\SiIV}. This
high--ionization phase has a solar metallicity. For comparison, the
system at z=$0.9276$ toward PG~$1206+459$ has a broad, smooth profile
for high--ionization transitions and is likely to be produced in a
spiral galaxy similar to Milky Way. On the other hand, the system at
z=$0.9254$ has resolved high--ionization, narrow components
superimposed on the low--ionization clouds, indicating that it is
likely to arise from a patchy distribution of gas in the outskirts of
the warped spiral disk \citep{1206}. In the case of this $z=0.7729$
system, due to the saturation of the {\CIV} absorption as well as lack
of the coverage of {\NV} in the spectrum, it is hard to rule out
either possibility.

The cloud at v$\sim225$~{\kms} is a typical satellite cloud of a
strong {\MgII} system (a weaker component found within hundreds of
{\kms} of the strong, dominant absorption).  Absorption separate from
that related to the dominant, low--ionization component, is also seen
in {\CIV} at similar velocity. This velocity overlap is found in most
satellite cloud systems at z$\sim1$ \citep{q1206}. However, with only
low--resolution {\it HST}/FOS spectra ($R=1,300$) available,
\citet{q1206} were not able to determine whether the offset {\CIV}
absorption was produced in a common halo encompassing the main galaxy
and the HVC or in kinematically distinct gas. The {\SiIV} absorption
in this system is stronger than what would be expected from an
ionizing source that is dominated by the extragalactic background
radiation, with the assumption of a solar abundance pattern. This
could be solved by a change in the abundance pattern, in particular,
an alpha--enhancement of $\sim0.7$ for this cloud. This would imply an
origin of the gas in Type II supernovae. If more satellite clouds at
z$\sim1$ are found to have a similar abundance pattern, then it could
suggest a more homogeneous origin of this type of HVC satellite at
this redshift, as opposed to a diverse production of the Milky Way
HVCs
\citep{sembachhvc}. On the other hand, an ionizing source with the
inclusion of significant ionizing radiation from stellar sources could
somewhat reduce the absorption in {\CII} and {\CIV} and hence reduce
the amount of increment needed for the alpha--elements. The relatively
blue, luminous galaxy, detected within an impact parameter of
$23.2~h^{-1}$~kpc from the absorber in the quasar field, could have
escaping radiation. The decrease in the extragalactic ionizing
background since z$\sim1$ could result in a general difference in the
relative absorption between the low-- and high--ionization transitions
in high--velocity clouds between the two epochs
\citep{hm01}.

\subsection{The $z=0.8545$ System Toward PG~$1248+401$}
\label{sec:z0.8545}

\subsubsection{Results}
\label{sec:r0.8545}

\begin{figure*}
\figurenum{5}
\epsscale{0.8}
\plotone{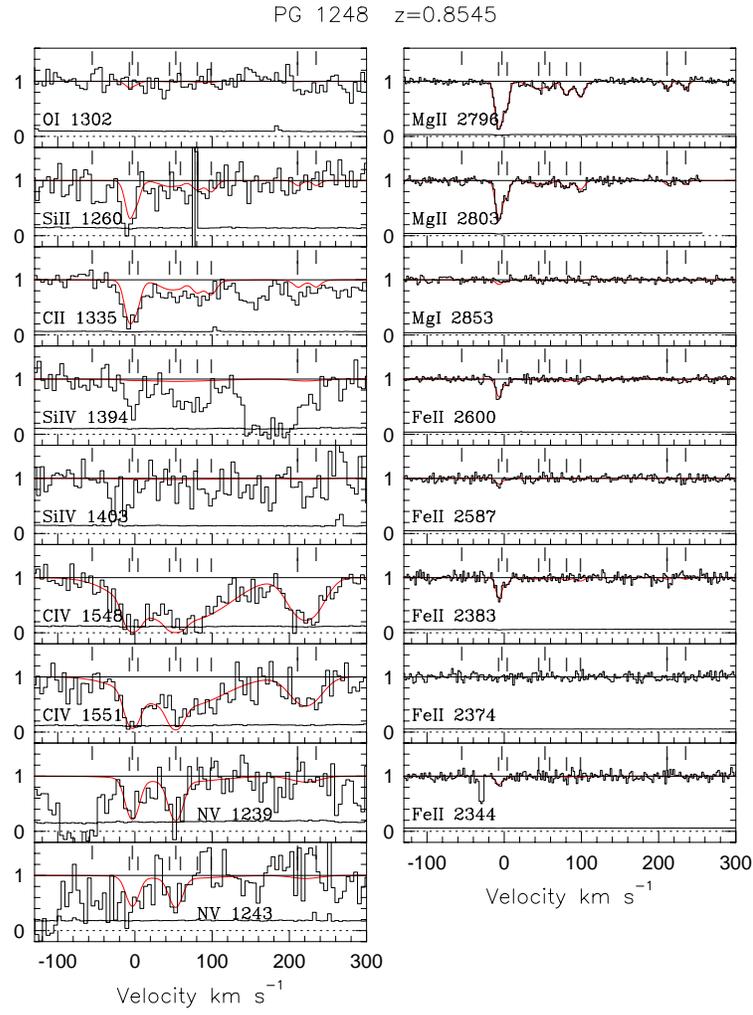}  
\protect\caption{
%\baselineskip = 0.7\baselineskip
The same as Fig~\ref{fig:0.8954}, except for the system at $z=0.8545$
toward PG~$1248+401$.}
\label{fig:0.8545}
\end{figure*}

As shown in Figure~\ref{fig:0.8545}, the system at $z=0.8545$ is a
multiple--cloud, weak {\MgII} system. The absorption in the {\MgII}
doublet spreads between $v\sim-20$ and $v\sim250$~{\kms} in
velocity space. At the system's velocity zero point, where the
strongest {\MgII} absorption occurs, absorption is also apparent in
various low-- and high--ionization transitions, such as {\FeII},
{\SiII}, {\CII}, {\SiIV}, {\CIV}, and {\NV}. The only exceptions are
{\MgI} and {\OI}, which do not have detected absorption in their
spectra.  In addition to having narrow components, aligned in velocity
space with low--ionization transitions, {\CIV} also appears to require
a broad, strong absorption feature, centered at v$\sim50$~{\kms}, to
fit its profiles.

Eight clouds were obtained from a simultaneous Voigt profile fit to
the spectra of {\MgII}, {\MgI}, and {\FeII}. Individual redshifts,
column densities, and Doppler parameters of these clouds are listed in
Table~\ref{tab:0.8545}. An extragalactic background, with radiation
from quasars only, was used as the incident spectrum. Assuming that
all eight clouds have the same metallicity, the metallicity was
initially set to one tenth solar value.  The ionization parameters of
the individual cloud are all constrained by $N({\FeII})$/$N({\MgII})$
to be in the range $-4.5 \le \log U \le -3.5$. At $\log U \ge -3.5$,
{\FeII} is underproduced, while at $\log U \le -4.5$, both {\FeII} and
{\OI} are overproduced.

Without coverage of the Lyman series lines, the system's metallicity
cannot be constrained. Both solar and one hundredth solar
metallicities have been explored in addition to the initial value of
one tenth solar metallicity. It follows that when the metallicity
decreases, the ionization parameter needs to increase and has a
narrower range. At $\log Z \sim -2$, the ionization parameter is
stringently constrained by $N({\FeII})$/$N({\MgII})$ to be $\log U
\sim -3.5$, within a $0.1$~dex of uncertainty. On the other hand, when
the metallicity increases to the solar value, the range for ionization
parameter is very similar to the one derived for $\log Z \sim
-1$. This is because when $\log Z \ge -1$, the clouds are optically
thin and the ratio of the column densities of metal transitions varies
only with ionization parameter. For $\log Z=-1$, cloud sizes range
from $0.5$--$6$~pc.

Regardless of the choice of metallicity, the derived ionization
parameter range gives rise to most of absorption seen in the
low--ionization transitions {\SiII} and {\CII}. {\CII}~$1335$ appears
to be contaminated by an known blend, since the absorption at
$\sim160$~{\kms} (see Figure~\ref{fig:0.8545}) is unmatched in any
other transitions. However, very little {\SiIV} and none of the {\CIV}
or {\NV} absorption could be produced in these clouds. Therefore, an
additional, more highly ionized phase is needed to give rise to these
high--ionization transitions.

Due to blending and noise in the spectra of {\SiIV}~$1394$ (blended
with Galactic {\FeII}~$2587$) and {\SiIV}~$1403$, a Voigt profile fit
was performed on the {\CIVdblt} doublet, as well as on the {\NVdblt}
doublet. Four clouds were obtained and their individual redshifts,
column densities, and Doppler parameters are listed in
Table~\ref{tab:0.8545}. In addition to the broad absorption profile,
the two clouds at $v\sim0$~{\kms} and $v\sim50$~{\kms} are narrow and
aligned with the low--ionization components in velocity space. Clearly
these fits are only approximate, but they still provide a guide to the
physical conditions of the high--ionization phase.

The high--ionization phase is assumed to have the same metallicity as
that of the {\MgII} phase. The two narrower clouds at $v\sim0$~{\kms}
and $v\sim50$~{\kms} cannot be produced by collisional ionization due
to their small Doppler parameters. A lower limit of $\log U \ge -1.8$
is placed on their ionization parameters so that they give rise to
enough {\NV} and do not overproduce {\SiIV} and {\CII} as well. For
the two broader clouds at $v\sim60$~{\kms} and $v\sim220$~{\kms}, both
photoionization and collisional ionization are explored. For the case
of photoionization, an upper limit of $\log U \le -1.5$ is placed on
the ionization parameter so that enough {\SiII} and {\CII} can be
produced. For the case of collisional ionization, the temperature is
constrained within $0.1$~dex of uncertainty of $\log T \sim 5.0$,
where {\CIV} peaks. At $\log T \ge 5.2$, {\NV} will be overproduced,
while at $\log T \le 4.8$, {\CII} and {\SiII} will be
overproduced. Since the clouds are optically thin, even at $\log Z
\sim -2$, the derived ionization parameter and temperature ranges are
consistent with metallicities within a reasonable range. For $\log Z =
-1$, the high--ionization cloud sizes are quite large, a few tens to a
few hundreds of kpc. They could be smaller for higher metallicity or
for alternative fits to the {CIV}.

If, alternatively, a contribution from star--forming galaxies is
included in the extragalactic background radiation, the ionization
parameter would need to be increased by $\sim0.5$~dex for the
high--ionization phase in order to fit all the transitions. 

\subsubsection{Discussion}
\label{sec:d0.8545}

The system at $z=0.8545$ is a multiple--cloud, weak {\MgII} absorber,
with low--ionization kinematics similar to that of many strong {\MgII}
systems.  In the formal multi--variate classification system of
\citet{archive2}, it was classified as a ``classic'' absorber.
As discussed in \S~\ref{sec:d0.5584}, such absorbers, with
one or more dominant {\MgII} clouds and several weaker ones spread
over a wider range in velocity space, are classified as
``kinematically spread'' systems. They are often found to have
physical conditions similar to those of classic, strong {\MgII}
absorbers such that they could arise in the outskirts of giant,
luminous galaxies. Other examples of ``kinematically spread''
multiple--cloud, weak {\MgII} absorbers include the systems at
$z=1.3250$ and $z=1.3430$ toward PG~$0117+213$.  In both of these
systems, the high--ionization gas, traced by {\OVI}, is found to be
kinematically distinct and offset from the low--ionization phase. In
fact, it is suggested in \citet{masiero} that this difference in
kinematics could be used to discriminate a difference in origin
between multiple--cloud, strong and multiple--cloud, weak absorbers,
with the latter arising in dwarf galaxies or in superwinds. However,
such kinematic distinction does not appear in the $z=0.5584$ system.
On the contrary, the high--ionization phase, traced by both narrow
structures and broad components in the {\CIV} absorption, is found to
be centered on the low--ionization {\MgII} clouds. In fact, with the
rest--frame equivalent width $W_{r}(2796$)$\sim0.25$~{\AA}, the system
at $z=0.8545$ is just below the cutoff for a strong {\MgII} absorber.
Thus, this system could be considered as a strong {\MgII} absorber,
only with somewhat weaker {\MgII} absorption. This ``blending of
classes'' should not be surprising since the $0.3$~{\AA} boundary
between strong and weak {\MgII} absorbers is only historical
\citep{ss92}.

The most interesting and striking feature of the low--ionization gas
in the $z=0.8545$ system is probably the three pairs of lines seen at
$v\sim45$ and $\sim59$~{\kms}, $v\sim81$ and $\sim99$~{\kms}, and
$v\sim210$ and $\sim235$~{\kms}, in the {\MgII} absorption. These
paired clouds, with $\delta$v$\sim14$, $\sim18$, and $\sim25$~{\kms},
respectively, could arise in the shell of a superbubble. Various
previous studies have provided evidence for a connection between
paired absorption lines and superbubbles.  For example, in
\citet{superbubble}, multiple pairs of lines, each with
$\delta$v$\sim30$~{\kms}, are detected in the $z=0.7443$ system toward
MC~$1331+170$. These line pairs are believed to arise in the
oppositely expanding sides of the cold superbubble shell which
intersects the line of sight.  Similarly, a pair of lines split by
$\sim20$~{\kms} is observed by \citet{ssh} in the Scutum
supershell. The kinematics of the {\CIV} in the $z=0.8545$ system are
consistent with the superbubble hypothesis as well.  Particularly for
the reddest pair, {\CIV} is centered in between the two {\MgII} clouds
in velocity space (at $v\sim220$~{\kms}) and has a broader ($b({\rm
C})\sim25$~{\kms}), unresolved absorption profile. Although not as
obvious, the other two pairs could have similar {\CIV} structures
within the uncertainties of our modeling of the high--ionization
phase.  High--ionization absorption could arise in the hotter, more
diffuse medium that exists in the interior of the shell which gives
rise to the paired {\MgII} absorption lines.

The kinematic structure of most strong {\MgII} systems is consistent
with a rotating disk in combination with radial infall from the halo
\citep{lb,kinmod,cv01}.  It is common for these absorbers to have
the high--ionization gas grouped with the low--ionization clouds.  In
the $z=0.8545$ system, the {\CIV} absorption, with both narrow
components and a broader, smooth structure in its profiles, is
centered on the {\MgII} clouds. The broad {\CIV} component, with a
Doppler parameter of $b({\rm C})\sim70$~{\kms}, could be comparable to
the high--ionization phase in system B of the double system at
$z\sim0.93$ toward PG~$1206+459$.  However, in contrast to the strong,
smooth {\NV} profiles in system B, which resemble the ``Galactic
corona'', the {\NV} absorption in the $z=0.8545$ system only has weak
and resolved structures. On the other hand, the high--ionization gas
in system A of the same $z\sim0.93$ double system has high--ionization
components centered on low--ionization groupings similar to those seen
in the {\CIV} and {\NV} profiles in the $z=0.8545$
system. Furthermore, like the system at $z=0.8545$, system A has
{\MgII} equivalent width $W_{r}(2796$)$\sim0.22$~{\AA}, just below the
threshold for strong {\MgII} absorption \citep{1206}.  Similar narrow
{\CIV} clouds are also present in the strong {\MgII} system at
$z=0.5504$ toward PG~$1241+706$, as described in
\S~\ref{sec:d0.5504}.  It is inferred that both system A toward 
PG~$1206+459$ and that system at $z=0.5504$ could arise in a patchy
distribution of gas in the outskirts of a giant galaxy (see
\S~\ref{sec:d0.5504}). 

Another system along the same line of sight toward PG~$1248+401$ is a
classic, strong {\MgII} absorber at $z=0.7729$. Similar to the
$z=0.8545$ system, it has very strong {\CIV} absorption. In addition,
it also has an offset satellite {\MgII} cloud at $v\sim220$~{\kms}
(though in this case, it is not paired absorption). As discussed in
\citet{archiveletter}, the physical conditions of the {\CIV} gas are
closely related to the presence of the kinematically outlying {\MgII}
clouds. In particular, the two could be governed by the same physical
processes. The balance between the outflow energetics from supernovae
in the galaxy and the galactic gravitational potential well could
result in a high--ionization Galactic--like ``corona'' in proportion
to the kinematics of gravitationally bound, cooling material
\citep{archiveletter}. The system at $z=0.8545$ is another good 
example in support of this conclusion.

In this case the most appealing comprehensive model would have
superbubbles giving rise to pairs of {\MgII} components with broad
surrounding {\CIV}. The superbubbles and the resulting satellite
clouds could also be related to production of a broader {\CIV}
``corona'' component. 

\subsection{The $z=0.6600$ System Toward PG~$1317+274$}
\label{sec:z0.6600}

\subsubsection{Results}
\label{sec:r0.6600}

As shown in Figure~\ref{fig:0.6600}, the system at $z=0.6600$ toward
PG~1317+274 (also known as CSO~873) is a multiple--cloud, strong
{\MgII} absorber. The absorption in {\MgI}, {\MgII}, and {\FeII}
spreads between $-50$ and $150$~{\kms} in velocity space and these
transitions all display a similar profile shape. The STIS spectrum has
coverage of the low--ionization tracers {\AlII}, {\SiII}~$1527$, and
the high--ionization tracers, {\SiIV} and {\CIV}. The {\SiIV} is not
detected in a noisy region of the spectra. Only a small amount of
{\CIV} absorption is detected at velocities consistent with the
{\MgII} clouds. However, strong {\CIV} absorption is detected at
v$\sim212$~{\kms}. The high--ionization transitions, {\NV} and {\OVI},
are not covered.

A simultaneous Voigt profile fit was performed on {\MgII}, {\MgI}, and
{\FeII}. Eight clouds were obtained, with their individual redshifts,
column densities, and Doppler parameters listed in
Table~\ref{tab:0.6600}. The clouds are assumed to all have the same
metallicity, with the value initially set to one tenth the solar
value. The ionization parameter of each cloud is constrained by
$N({\FeII})$/$N({\MgII})$, wherever {\FeII} is detected.  The clouds
at $v\sim-10$, $0$, $12$, $90$, and $100$~{\kms} have detected {\FeII}
and their ionization parameters are roughly determined to be $-7 \le
\log U \le -5$. Within this ionization parameter range, the observed
$N({\MgI})$/$N({\MgII})$ can be produced as well. At $v\sim50$, $78$,
and $140$~{\kms}, neither {\MgI} nor {\FeII} is detected. Therefore,
only very limited constraints can be placed on their ionization
parameters. For simplicity, we assume the clouds at these velocities
to have ionization parameters in a similar range, $-7
\le \log U \le -5$, as the others. 

Lacking the coverage of the Lyman series lines, metallicity cannot be
constrained. In addition to our initial assumption of one tenth solar
metallicity, both solar and one hundredth solar metallicities have
been considered. We found a general trend that when the metallicity
increased, a slightly higher ionization parameter was required to give
rise to the observed {\FeII} and {\MgI}; and when the metallicity
decreased, a lower ionization parameter was needed instead. However,
this adjustment is usually within $0.1$ or $0.2$~dex. Therefore, the
ionization ranges that are derived using a tenth solar metallicity are
applicable for a wide metallicity range.

Because $\log U \le -5$, based upon detected {\FeII} absorption,
{\CIV} is not significantly produced by the {\MgII}
clouds. Furthermore, there is an additional {\CIV} component at
$v\sim212$~{\kms}. Therefore, a high--ionization phase is needed.  A
Voigt profile fit to the {\CIVdblt} yielded four components (see
Table~\ref{tab:0.6600}). We assume the same metallicity, $\log Z
\simeq -1$, as we used for the low--ionization phases. The ionization 
parameters were constrained by the low--ionization transitions. A
lower limit of $\log U \ge -2.5$ applies for the clouds at $v\sim-8$,
$79$, and $153$~{\kms} in order that {\SiII} and {\AlII} are not
overproduced. Constrained by the same transitions, $\log U \ge -2$ for
the cloud at $212$~{\kms}. 

An alternative input spectrum, with the inclusion of the stellar
contribution escaping from galaxies, was also superimposed on the test
models. It has been shown that there is little difference in the
derived ranges for the ionization parameter. This is because at the
energies relevant to photoionization of the low--ionization
transitions, the two input spectra have very similar shapes. For the
high--ionization phase, the limits on $\log U$ would be somewhat
larger due to the change in input spectral shape.

\subsubsection{Discussion}
\label{sec:d0.6600}

The system at z=$0.6600$ toward PG~$1317+274$ is a strong {\MgII}
absorber that has been classified as ``{\CIV}--deficient''
because its {\CIV} absorption is weak relative to {\MgII}
\citep{archive2}.  In fact, there are only small amounts of {\CIV} absorption
detected in the STIS spectrum at the velocities of the {\MgII}
components.  However, there is a strong ``{\CIV} high--velocity cloud''
at $v\sim212$~{\kms}.  All of the {\CIV} clouds required a separate
high--ionization phase, with $\log U \sim -2$, since the low--ionization
clouds were constrained to have $\log U < -5$ in order that they would
produce the detected {\FeII} absorption.

A galaxy is found within an impact parameter of $71.6~h^{-1}$~kpc of
the line of sight toward PG~$1317+276$ in the {\it HST}/WFPC2 F702W
image of the quasar field (see Table~\ref{tab:galaxy}.  This galaxy is
at a redshift consistent with that of the absorber.  It is a reddish
spiral ($<$B-K$>=3.84$ \citep{csv96}) with a rotation curve that
extrapolates to the velocity of the observed {\MgII} absorption
\citep{steidel02}.  Because of the large impact parameter, even in projection, this
would imply what is perhaps an unrealistically large rotating disk,
leading \citet{steidel02} to the conclusion that the halo is rotating.
Also, due to its large impact parameter, its
identification as the galaxy responsible for the absorption is
somewhat tentative, though it seems that then a dwarf companion galaxy
would be responsible.  It may be an important clue to the nature of
this system that the strong {\CIV} absorption component is at a
velocity only slightly redward ($\sim 50$~{\kms}) of the galaxy
systemic velocity \citep{cwcboulder}.

The system at z=$0.9902$ toward PG~$1634+706$ is also a
{\CIV}--deficient system. In that case the relatively weak {\CIV} absorption is
mostly produced in the low--ionization phase which gives rise to
{\MgII} \citep{1634}. The diffuse phase in the system, required by the
broad {\Lya} absorption profile, has a very low metallicity (less than
$0.01$ solar). Other {\CIV}--deficient absorbers include the
z=$0.7290$ and z=$1.0480$ systems toward PG~$0117+213$
\citep{masiero}.  In both of those systems, {\CIV} absorption was
not even detected in STIS spectra.  In none of these three {\CIV}--deficient
systems was an additional high--ionization phase required.  
Our $z=0.6600$ system toward PG~$1317+274$, at first glance, differs
since even the {\CIV} at the same velocities as {\MgII} requires
a separate phase.  However, it is also possible that there is a
very--low ionization phase, as we discussed in \S~\ref{sec:r0.5504}
for the case of the $z=0.5504$ absorber toward PG~$1241+176$ and in
\citet{1634} for the $z=0.9902$ absorber toward PG~$1634+706$.
In this case, the {\FeII} absorption could arise from the very--low
ionization phase and the ionization parameter of the {\MgII} clouds
could be higher in order to produce some {\CIV} absorption.  Higher
resolution data would be required to distinguish these cases.
In any case, for none of these {\CIV}--deficient systems is there
a strong and/or broad high--ionization component indicative of a
corona.

It seems that {\CIV}--deficient systems could have a various types of
galaxy hosts, including elliptical galaxies, quiescent disk galaxies,
and even star--forming disk galaxies. The key factor is that the
coronal phase that gives rise to the {\CIV} absorption is either
absent or too highly ionized.  In the case of this $z=0.6600$ absorber,
we hypothesize that the corona is absent at the high impact parameter
in a relatively red spiral.  This explanation also requires the
unusual circumstance of passing through a dense region of a rotating
disk or rotating halo that would produce strong (and not just weak)
{\MgII} absorption.

For the case of this $z=0.6600$ systems, the $v=212$~{\kms} can be
viewed somewhat independently of the {\CIV}--deficient nature of the
system.  It is quite reminiscent of the ``{\CIV} high--velocity clouds''
found by \citet{sembach95} along the Galactic line of sight toward
Markarian 509.  It does not seem unusual that such a high velocity
cloud could be within $50$~{\kms} of the galaxy systemic velocity.
It is this high velocity component that gives this system relatively
strong {\CIV} absorption for a {\CIV}--deficient system.

\section{Summary and Conclusion}
\label{sec:conclusion}

In this paper we present the modeling results of six {\MgII}--selected
absorption systems, in the redshift interval of $0.5$ to $0.9$, along
the lines of sight toward three quasars.  These six absorption systems
sample a variety of types of absorbers and are therefore
representative of a larger population in this redshift interval. Due
to lack of coverage of the Lyman series lines in five of the systems,
we were unable to obtain constraints on metallicities for them.
However, we were able to constrain the number of phases of gas along
the line of sight and their densities and kinematic structures. Here
we summarize these results and their implications:

The system at $z=0.8954$ toward PG~$1241+176$ is a single--cloud, weak
{\MgII} absorber. This is the only system in this paper that has
spectral coverage of {\Lya}. Due to the slight offset of the {\MgII}
absorption from the other transitions, this system could arise in
either a one--phase or a two--phase model. Although the {\CIV}
absorption is produced by gas that has a $\sim2\%$ solar metallicity,
lower than what is usually found in single--cloud, weak {\MgII}
absorbers \citep{weak2,weak1634}, the {\MgII} absorption in this
system could still arise in gas with a near--solar metallicity for the
case of a two--phase scenario. Alternatively, this system could simply
have a trace of {\MgII} absorption detected from a lone {\CIV} cloud,
a situation that would be more common at lower redshift due to a
decreasing extragalactic background radiation \citep{anand}.

The system at $z=0.5584$ toward PG~$1241+176$ is a kinematically
compact, multiple--cloud, weak {\MgII} absorber. It has been suggested
that systems in this classification arise in dwarf galaxies based upon
low metallicity and distinctive kinematic offsets ($\sim50$~{\kms})
between low-- and high--ionization components
\citep{zonak}.  With spectral coverage of only limited transitions
({\MgII}, {\AlII}, {\AlIII}, and {\CIV}), we cannot constrain the
physical properties of this system too well.  {\CIV} can arise in the
same phase as the {\MgII}, but in the absence of coverage of {\OVI},
the presence of ``offset phases'' cannot be considered. In the absence
of {\HI} coverage, neither can metallicity, though properties of the
low--ionization phases are similar between this system and the
$z=1.0414$ system toward PG~$1634+706$ \citep{zonak}. An answer to the
question of whether kinematically compact, multiple--cloud, weak
{\MgII} absorption is a distinctive signature of dwarfs or whether it
arises in a variety of circumstances awaits data on additional systems
with large spectral coverage.

The system at $z=0.8545$ toward PG~$1248+401$ is a kinematically
spread, multiple--cloud, weak {\MgII} absorber. It has absorption
features similar to those seen in classic strong {\MgII} absorbers.
Three pairs of {\MgII} absorption features are seen in the profiles,
suggesting that this system may arise in the cold shells of a series
of expanding superbubbles \citep{superbubble}. Since there has been no
host galaxy identification, we suggest that this system could arise in
the outskirts of a luminous galaxy. A broad high--ionization phase,
consistent with a corona, is also apparent.

The systems at $z=0.5504$ toward PG~$1241+176$ and at $z=0.7729$
toward PG~$1248+401$ are both classic strong {\MgII} absorbers. Like
other classic strong {\MgII} absorbers, both low-- and
high--ionization phases, with some kinematic association with each
other, are needed to account for absorption seen in various
transitions. Strong {\MgII} absorption systems at $z\sim1$ are known
to be associated with luminous galaxies with a variety of morphology
types \citep{s95,csv96,steidel02}. While the resolved structures in
the high--ionization transitions in the system at $z=0.5504$ toward
PG~$1241+176$ indicate a patchy corona, the saturated {\CIV}
absorption in the system at $z=0.7729$ toward PG~$1248+401$ could
arise in a traditional Milky Way--like corona. Both systems have
identified galaxy hosts in their vicinity.

The system at $z=0.6600$ toward PG~$1317+274$ is a {\CIV}--deficient
absorber. There are multiple causes for a system to be
{\CIV}--deficient, for example, an early--type galaxy host
\citep{1634}, a low--metallicity corona \citep{archive2}, or a corona
that is too highly ionized. The identification of a spiral galaxy
at an impact parameter of $71.6h^{-1}$~kpc host suggests that the
``{\CIV}--deficiency'' in this system could result from suppressed or
patchy coronal structure at a large impact parameter in the host
galaxy.  The strongest {\CIV} absorption component for this system
is at a velocity of $v\sim212$~{\kms} and is not coincident with
the {\MgII} absorption.  We propose that this component, which
is at close to the systemic velocity of the candidate absorbing
galaxy, is an analog to the Milky Way ``{\CIV} high--velocity clouds''
\citep{sembach95}.

Through this detailed study of these six systems, we have learned
of the varied relationships between low and high--ionization gas
in {\MgII} absorption line systems.  We have discussed the implications
with respect to the absorption signatures of galactic coronae.  We
have also added examples of single--cloud and multiple--cloud, weak
{\MgII} absorbers to the handful of such systems studied at this
level of detail.  The next step in reaching a global understanding
of absorption properties is to combine the modeling results of this
study, with those on other systems that were described in previous papers
{1634,1206,weak1634,zonak,masiero}.  Using inferred physical properties,
it should be possible to make quantitative comparisons of the gas properties
across the range of gaseous structures sampled by {\MgII} absorption
at intermediate redshifts.

This research was funded by NASA under grants NAG~$5$--$6399$,
NNG$04$GE$73$G, and {\it HST}--GO--$08672.01$--A, the latter from the
Space Telescope Science Institute, which is operated by AURA, Inc.,
under NASA contract NAS~$5$--$26555$; and by NSF under grant
AST--$04$--$07138$. JRM was partially funded by the NSF REU program.

%%%%%%%%%%%%%%%%%%%%%%%%%%%%%%%%%%%%%%%%%%%%%%%%%%%%%%%%%%%%%%%%%%%%%%%%%%%%%%%%%%

%%%%%%%%%%%%%%%%%%%%%%%%%%%%%%%%%%%%%%%%%%%%%%%%%%%%%%

\clearpage

\newpage

\newpage

\newpage

\newpage

\newpage

%%%%%%%%%%%%%%%%%%%% TABLES %%%%%%%%%%%%%%%%%%%%
%%%%%%%%%%%%%%
%%%%%%%%%%%%%%
%%%%%%%%%%%%%%

%
%EW tables first
%
\clearpage

\begin{deluxetable}{lcccccc}
\tablenum{1}
\tabletypesize{\footnotesize}
\rotate
\tablewidth{0pt}
\tablecaption{Spectroscopic Observation}
\tablehead{
\colhead{Quasar}   &
\colhead{Date} &
\colhead{Wavelength}   &
\colhead{Exposure} &
\colhead{Date} &
\colhead{Wavelength}   &
\colhead{Exposure} \\
\colhead{}   &
\colhead{} &
\colhead{\AA}   &
\colhead{s} &
\colhead{} &
\colhead{\AA}   &
\colhead{s}
}

\startdata
PG$~1241+176$ & June $2002$ & $2270$--$3100$ & $19221$ & January $1995$ & $3765$--$6190$ & $2400$\\
PG$~1248+401$ & July $2001$ & $2270$--$3100$ & $25206$ & January $1995$ & $3765$--$6190$ & $4200$\\
PG$~1317+274$ & January $2001$; June $2002$ & $2270$--$3100$ & $5317$; $8240$ & January $1995$ & $3810$--$6315$ & $3600$\\

\hline
\enddata
\vglue -0.05in
\tablecomments{
Detailed observing information is listed in the table. Ultraviolet and
optical spectra are obtained with {\it HST}/STIS ($R=30,000$) and
Keck/HIRES ($R=45,000$), respectively. The date, wavelength coverage, and total
exposure time of {\it HST}/STIS observation are listed in columns $2$,
$3$, $4$. The Keck/HIRES information is listed in columns $5$, $6$,
$7$.}
\label{tab:spectra}
\end{deluxetable}
\clearpage

\newpage
\begin{deluxetable}{llccccc}
\tablenum{2}
\tabletypesize{\footnotesize}
%\rotate
\tablewidth{0pt}
\tablecaption{Galaxy Properties}
\tablehead{
\colhead{Quasar} &
\colhead{$z_{abs}$}   &
\colhead{$L_B$}   &
\colhead{$L_K$}   &
\colhead{B-K}   &
\colhead{Impact Parameter} &
\colhead{Morphology}\\  
\colhead{} &
\colhead{}   &
\colhead{[$L_B^*$]}   &
\colhead{[$L_K^*$]}   &
\colhead{} &
\colhead{[$h^{-1}$~kpc]} &
\colhead{}
}

\startdata
PG~$1241+176$ & $0.5504$ & $0.64$ & $0.41$ & $3.42$ & $13.8$ & \nodata \\
PG~$1248+401$ & $0.7729$ & $0.53$ & $0.27$ & $3.15$ & $23.2$ & \nodata \\
PG~$1317+274$ & $0.6600$ & $1.51$ & $1.90$ & $3.84$ & $71.6$ & spiral \\

\hline
\enddata
\vglue -0.05in
\tablecomments{
\baselineskip=0.7\baselineskip
The B--band and K--band luminosities, rest--frame $<$B-K$>$ color, 
and the impact parameter are listed for individual galaxies that are
detected in the quasar field, along with the redshift of the 
corresponding absorption systems.  All galaxy information is taken
from \citet{csv96}, except for morphology which is taken from
\citet{steidel02}.}
\label{tab:galaxy}
\end{deluxetable}
\clearpage

\newpage
\begin{deluxetable}{lccccc}
\tablenum{3}
\tabletypesize{\footnotesize}
%\rotate
\tablewidth{0pt}
\tablecaption{Rest Frame Equivalent Widths}
\tablehead{
\colhead{Quasar}   &
\colhead{System}   &
\colhead{\MgII~$2796$ (\AA)}   &
\colhead{\MgI~$2853$ (\AA)}   &
\colhead{\FeII~$2600$ (\AA)}   &
\colhead{\CIV~$1548$ (\AA)} 
}

\startdata
PG$~1241+176$ & $z=0.8954$ & $0.018\pm0.005$ & $<0.004$ & $<0.005$ & $0.104\pm0.016$\\
$$ & $z=0.5584$ & $0.135\pm0.014$ & $<0.008$ & $<0.015$ & $0.251\pm0.021$\\
$$ & $z=0.5504$ & $0.481\pm0.019$ & $0.098\pm0.030$ & $0.236\pm0.048$ & $0.890\pm0.026$\\
PG$~1248+401$ & $z=0.7729$ & $0.694\pm0.009$ & $0.065\pm0.022$ & $0.247\pm0.020$ & $0.785\pm0.031$\\
$$ & $z=0.8545$ & $0.253\pm0.014$ & $<0.019$ & $0.031\pm0.007$ & $0.871\pm0.024$\\
PG$~1317+274$ & $z=0.6600$ & $0.338\pm0.011$ & $0.026\pm0.009$ & $0.126\pm0.016$ & $0.316\pm0.022$\\

\hline
\enddata
\vglue -0.05in
\tablecomments{
\baselineskip=0.7\baselineskip
This table lists the rest--frame equivalent widths for {\MgII}~$2796$,
{\MgI}~$2853$, {\FeII}~$2600$, and {\CIV}~$1548$ in all 6 systems,
with the $\sigma$ value for each measurement included. If the
transition is not detected, then a $3\sigma$ level limit is
given.}
\label{tab:ew}
\end{deluxetable}
\clearpage

\newpage
\begin{deluxetable}{cccccccccccc}
\tablenum{4}
\tabletypesize{\scriptsize}
\rotate
\tablewidth{0pt}
\tablecaption{Model Parameters for the $z=0.8954$ System Toward PG~$1241+176$}
\tablehead{
\colhead{Cloud}   &
\colhead{Optimized}   &
\colhead{Velocity}   &
\colhead{$\log N$}   &
\colhead{b} &
\colhead{$\log N_{tot}$}   &
\colhead{$\log N ({\HI})$}   &
\colhead{Z}   &
\colhead{$\log U$} &
\colhead{T} &
\colhead{Size} &
\colhead{Density} \\
\colhead{Number}   &
\colhead{Transition}   &
\colhead{[{\kms}]}   &
\colhead{[{\cmsq}]}   &
\colhead{[{\kms}]} &
\colhead{[Z$_{\sun}$]}   &
\colhead{[{\cmsq}]}   &
\colhead{[{\cmsq}]}   &
\colhead{} &
\colhead{[K]} &
\colhead{[kpc]} &
\colhead{[{\cc}]}
}

\startdata
\multicolumn{12}{c}{One--Phase Model}\\
\hline
$1$ & {\CIV} & $-5$ & $14.08$ & $6.7$ & $20.2$ & $16.7$ & $0.02$ & $-2.1$ & $26000$ & $100$ & $0.0005$ \\
\hline
\multicolumn{12}{c}{Two--Phase Model}\\
\hline
$1$ & {\MgII} & $0$ & $11.7$ & $6.8$ & $16.3$ & $15.2$ & $1.0$ & $-4.0$ & $6500$ & $0.2$ & $0.04$ \\
$1$ & {\CIV} & $-5$ & $14.08$ & $6.7$ & $20.0$ & $16.2$ & $0.02$ & $-1.7$ & $35000$ & $190$ & $0.0002$ \\

\hline
\enddata
\vglue -0.05in
\tablecomments{
\baselineskip=0.7\baselineskip
Model parameters for the system at $z=0.8954$. A fit of this model
is superimposed on the data in Fig.~\ref{fig:0.8954}. Column densities
are listed in logarithmic units.
}
\label{tab:0.8954}
\end{deluxetable}
\clearpage

\newpage
\begin{deluxetable}{cccccccccccc}
\tablenum{5}
\tabletypesize{\scriptsize}
\rotate
\tablewidth{0pt}
\tablecaption{Model Parameters for the $z=0.5584$ System Toward PG~$1241+176$}
\tablehead{
\colhead{Cloud}   &
\colhead{Optimized}   &
\colhead{Velocity}   &
\colhead{$\log N$}   &
\colhead{b} &
\colhead{$\log N_{tot}$}   &
\colhead{$\log N ({\HI})$}   &
\colhead{Z}   &
\colhead{$\log U$} &
\colhead{T} &
\colhead{Size} &
\colhead{Density} \\
\colhead{Number}   &
\colhead{Transition}   &
\colhead{[{\kms}]}   &
\colhead{[{\cmsq}]}   &
\colhead{[{\kms}]} &
\colhead{[{\cmsq}]}   &
\colhead{[{\cmsq}]}   &
\colhead{[Z$_{\sun}$]}   &
\colhead{} &
\colhead{[K]} &
\colhead{[kpc]} &
\colhead{[{\cc}]}
}

\startdata
$1$ & {\MgII} & $-25$ & $11.89$ & $2.7$ & $19.3$ & $16.2$ & $0.1$ & $-2.3$ & $20000$ & $16$ & $0.0004$ \\
$2$ & {\MgII} & $-8$ & $12.05$ & $8.5$ & $19.4$ & $16.3$ & $0.1$ & $-2.3$ & $20000$ & $23$ & $0.0004$ \\
$3$ & {\MgII} & $9$ & $11.98$ & $3.0$ & $19.2$ & $16.2$ & $0.1$ & $-2.4$ & $20000$ & $11$ & $0.0005$ \\
$4$ & {\MgII} & $19$ & $11.91$ & $3.2$ & $18.6$ & $16.1$ & $0.1$ & $-2.8$ & $16000$ & $1$ & $0.001$ \\

\hline
\enddata
\vglue -0.05in
\tablecomments{
\baselineskip=0.7\baselineskip
The same as Table~\ref{tab:0.8954}, except for the system at $z=0.5584$.
}
\label{tab:0.5584}
\end{deluxetable}
\clearpage

\newpage
\begin{deluxetable}{cccccccccccc}
\tablenum{6}
\tabletypesize{\scriptsize}
\rotate
\tablewidth{0pt}
\tablecaption{Model Parameters for the $z=0.5504$ System Toward PG~$1241+176$}
\tablehead{
\colhead{Cloud}   &
\colhead{Optimized}   &
\colhead{Velocity}   &
\colhead{$\log N$}   &
\colhead{b} &
\colhead{$\log N_{tot}$}   &
\colhead{$\log N ({\HI})$}   &
\colhead{Z}   &
\colhead{$\log U$} &
\colhead{T} &
\colhead{Size} &
\colhead{Density} \\
\colhead{Number}   &
\colhead{Transition}   &
\colhead{[{\kms}]}   &
\colhead{[{\cmsq}]}   &
\colhead{[{\kms}]} &
\colhead{[{\cmsq}]}   &
\colhead{[{\cmsq}]}   &
\colhead{[Z$_{\sun}$]}   &
\colhead{} &
\colhead{[K]} &
\colhead{[kpc]} &
\colhead{[{\cc}]}
}

\startdata
$1$ & {\MgII} & $-8$ & $12.31$ & $8.1$ & $18.5$ & $18.4$ & $0.1$ & $-6$ & $5600$ & $0.0006$ & $2$ \\
$2$ & {\MgII} & $2$ & $13.07$ & $19.4$ & $18.9$ & $18.9$ & $0.1$ & $-6$ & $210$ & $0.002$ & $2$ \\
$3$ & {\MgII} & $40$ & $13.52$ & $5.9$ & $17.8$ & $17.6$ & $0.1$ & $-6$ & $7800$ & $0.0001$ & $2$ \\
$4$ & {\MgII} & $145$ & $11.96$ & $6.5$ & $17.4$ & $17.2$ & $0.1$ & $-6$ & $7900$ & $0.00004$ & $2$ \\
$1$ & {\CIV} & $-51$ & $14.13$ & $36.0$ & $19.3$ & $15.9$ & $0.1$ & $-2$ & $24000$ & $37$ & $0.0002$ \\
$2$ & {\CIV} & $16$ & $13.87$ & $25.7$ & $19.1$ & $15.6$ & $0.1$ & $-2$ & $23000$ & $20$ & $0.0002$ \\
$3$ & {\CIV} & $65$ & $13.50$ & $11.9$ & $18.7$ & $15.2$ & $0.1$ & $-2$ & $24000$ & $8$ & $0.0002$ \\
$4$ & {\CIV} & $103$ & $13.72$ & $10.6$ & $18.9$ & $15.4$ & $0.1$ & $-2$ & $24000$ & $14$ & $0.0002$ \\
$5$ & {\CIV} & $149$ & $13.81$ & $12.4$ & $19.0$ & $15.5$ & $0.1$ & $-2$ & $24000$ & $18$ & $0.0002$ \\
$6$ & {\CIV} & $173$ & $13.73$ & $13.7$ & $18.9$ & $15.5$ & $0.1$ & $-2$ & $24000$ & $15$ & $0.0002$ \\

\hline
\enddata
\vglue -0.05in
\tablecomments{
\baselineskip=0.7\baselineskip
The same as Table~\ref{tab:0.8954}, except for the system at $z=0.5504$.
An abundance pattern adjustment of {\rm Ca} by $-0.5$~dex has been applied
to {\MgII} cloud $3$.}
\label{tab:0.5504}
\end{deluxetable}
\clearpage

\newpage
\begin{deluxetable}{cccccccccccc}
\tablenum{7}
\tabletypesize{\scriptsize}
\rotate
\tablewidth{0pt}
\tablecaption{Model Parameters for the $z=0.7729$ System Toward PG~$1248+401$}
\tablehead{
\colhead{Cloud}   &
\colhead{Optimized}   &
\colhead{Velocity}   &
\colhead{$\log N$}   &
\colhead{b} &
\colhead{$\log N_{tot}$}   &
\colhead{$\log N ({\HI})$}   &
\colhead{Z}   &
\colhead{$\log U$} &
\colhead{T} &
\colhead{Size} &
\colhead{Density} \\
\colhead{Number}   &
\colhead{Transition}   &
\colhead{[{\kms}]}   &
\colhead{[{\cmsq}]}   &
\colhead{[{\kms}]} &
\colhead{[{\cmsq}]}   &
\colhead{[{\cmsq}]}   &
\colhead{[Z$_{\sun}$]}   &
\colhead{} &
\colhead{[K]} &
\colhead{[kpc]} &
\colhead{[{\cc}]}
}
\startdata
$1$ & {\MgII} & $-30$ & $12.66$ & $8.5$ & $18.3$ & $17.0$ & $0.1$ & $-4$ & $11000$ & $0.02$ & $0.03$ \\
$2$ & {\MgII} & $-13$ & $12.80$ & $5.7$ & $18.3$ & $17.9$ & $0.1$ & $-5$ & $9200$ & $0.002$ & $0.3$ \\
$3$ & {\MgII} & $-5$ & $13.33$ & $7.4$ & $18.8$ & $18.7$ & $0.1$ & $-5.5$ & $6100$ & $0.002$ & $0.9$ \\
$4$ & {\MgII} & $13$ & $12.31$ & $3.5$ & $18.0$ & $16.7$ & $0.1$ & $-4$ & $11000$ & $0.01$ & $0.03$ \\
$5$ & {\MgII} & $24$ & $11.70$ & $1.4$ & $18.2$ & $15.8$ & $0.1$ & $-3$ & $15000$ & $0.2$ & $0.003$ \\
$6$ & {\MgII} & $40$ & $13.26$ & $6.5$ & $18.9$ & $17.7$ & $0.1$ & $-4$ & $11000$ & $0.09$ & $0.03$ \\
$7$ & {\MgII} & $52$ & $11.89$ & $5.3$ & $17.6$ & $16.3$ & $0.1$ & $-4$ & $12000$ & $0.004$ & $0.03$ \\
$8$ & {\MgII} & $225$ & $12.30$ & $5.5$ & $18.0$ & $16.7$ & $0.1$ & $-4$ & $11000$ & $0.01$ & $0.03$ \\
$1$ & {\SiIV} & $-57$ & $12.86$ & $9.0$ & $18.2$ & $14.9$ & $1$ & $-2$ & $11000$ & $1.6$ & $0.0003$ \\
$2$ & {\SiIV} & $-14$ & $13.78$ & $27.0$ & $19.1$ & $15.8$ & $1$ & $-2$ & $11000$ & $13$ & $0.0003$ \\
$3$ & {\SiIV} & $36$ & $13.56$ & $14.3$ & $18.8$ & $15.6$ & $1$ & $-2$ & $11000$ & $8$ & $0.0003$ \\
$4$ & {\SiIV} & $193$ & $12.05$ & $3.3$ & $17.3$ & $14.1$ & $1$ & $-2$ & $12000$ & $0.2$ & $0.0003$ \\
$5$ & {\SiIV} & $224$ & $12.80$ & $9.1$ & $18.1$ & $14.9$ & $1$ & $-2$ & $12000$ & $1$ & $0.0003$ \\

\hline
\enddata
\vglue -0.05in
\tablecomments{
\baselineskip=0.7\baselineskip
The same as Table~\ref{tab:0.8954}, except for the system at $z=0.7729$.
An $\alpha$--enhancement of $0.7$~dex was applied to {\SiIV} clouds $4$ 
and $5$.}
\label{tab:0.7729}
\end{deluxetable}
\clearpage

\newpage
\begin{deluxetable}{cccccccccccc}
\tablenum{8}
\tabletypesize{\scriptsize}
\rotate
\tablewidth{0pt}
\tablecaption{Model Parameters for the $z=0.8545$ System Toward PG~$1248+401$}
\tablehead{
\colhead{Cloud}   &
\colhead{Optimized}   &
\colhead{Velocity}   &
\colhead{$\log N$}   &
\colhead{b} &
\colhead{$\log N_{tot}$}   &
\colhead{$\log N ({\HI})$}   &
\colhead{Z}   &
\colhead{$\log U$} &
\colhead{T} &
\colhead{Size} &
\colhead{Density} \\
\colhead{Number}   &
\colhead{Transition}   &
\colhead{[{\kms}]}   &
\colhead{[{\cmsq}]}   &
\colhead{[{\kms}]} &
\colhead{[{\cmsq}]}   &
\colhead{[{\cmsq}]}   &
\colhead{[Z$_{\sun}$]}   &
\colhead{} &
\colhead{[K]} &
\colhead{[kpc]} &
\colhead{[{\cc}]}
}

\startdata
$1$ & {\MgII} & $-7$ & $12.77$ & $4.3$ & $18.4$ & $17.1$ & $0.1$ & $-4$ & $11000$ & $0.03$ & $0.03$ \\
$2$ & {\MgII} & $4$ & $12.10$ & $2.2$ & $17.8$ & $16.5$ & $0.1$ & $-4$ & $12000$ & $0.006$ & $0.03$ \\
$3$ & {\MgII} & $45$ & $11.95$ & $16.3$ & $17.6$ & $16.3$ & $0.1$ & $-4$ & $12000$ & $0.004$ & $0.03$ \\
$4$ & {\MgII} & $59$ & $11.02$ & $1.6$ & $16.7$ & $15.4$ & $0.1$ & $-4$ & $12000$ & $0.0005$ & $0.03$ \\
$5$ & {\MgII} & $81$ & $11.88$ & $5.7$ & $17.6$ & $16.3$ & $0.1$ & $-4$ & $12000$ & $0.003$ & $0.03$ \\
$6$ & {\MgII} & $99$ & $11.98$ & $5.9$ & $17.7$ & $16.4$ & $0.1$ & $-4$ & $12000$ & $0.004$ & $0.03$ \\
$7$ & {\MgII} & $211$ & $11.50$ & $3.3$ & $17.2$ & $15.9$ & $0.1$ & $-4$ & $12000$ & $0.001$ & $0.03$ \\
$8$ & {\MgII} & $235$ & $11.49$ & $3.4$ & $17.2$ & $15.9$ & $0.1$ & $-4$ & $12000$ & $0.001$ & $0.03$ \\
$1$ & {\CIV} & $-3$ & $14.14$ & $8.8$ & $19.6$ & $15.3$ & $0.1$ & $-1.2$ & $34000$ & $245$ & $0.00005$ \\
$2$ & {\CIV} & $55$ & $14.41$ & $67.8$ & $19.6$ & $15.9$ & $0.1$ & $-1.8$ & $26000$ & $55$ & $0.0002$ \\
$3$ & {\CIV} & $221$ & $13.93$ & $24.6$ & $19.1$ & $15.4$ & $0.1$ & $-1.8$ & $25000$ & $18$ & $0.0002$ \\
$1$ & {\NV} & $53$ & $13.81$ & $9.0$ & $19.6$ & $15.3$ & $0.1$ & $-1.2$ & $32000$ & $260$ & $0.00005$ \\

\hline
\enddata
\vglue -0.05in
\tablecomments{
\baselineskip=0.7\baselineskip
The same as Table~\ref{tab:0.8954}, except for the system at $z=0.8545$.
}
\label{tab:0.8545}
\end{deluxetable}
\clearpage

\newpage
\begin{deluxetable}{cccccccccccc}
\tablenum{9}
\tabletypesize{\scriptsize}
\rotate
\tablewidth{0pt}
\tablecaption{Model Parameters for the $z=0.6600$ System Toward PG~$1317+274$}
\tablehead{
\colhead{Cloud}   &
\colhead{Optimized}   &
\colhead{Velocity}   &
\colhead{$\log N$}   &
\colhead{b} &
\colhead{$\log N_{tot}$}   &
\colhead{$\log N ({\HI})$}   &
\colhead{Z}   &
\colhead{$\log U$} &
\colhead{T} &
\colhead{Size} &
\colhead{Density} \\
\colhead{Number}   &
\colhead{Transition}   &
\colhead{[{\kms}]}   &
\colhead{[{\cmsq}]}   &
\colhead{[{\kms}]} &
\colhead{[{\cmsq}]}   &
\colhead{[{\cmsq}]}   &
\colhead{[Z$_{\sun}$]}   &
\colhead{} &
\colhead{[K]} &
\colhead{[kpc]} &
\colhead{[{\cc}]}
}

\startdata
$1$ & {\MgII} & $-10$ & $12.63$ & $7.3$ & $18.1$ & $18.0$ & $0.1$ & $-6$ & $7600$ & $0.0002$ & $2$ \\
$2$ & {\MgII} & $0$ & $12.69$ & $4.6$ & $18.1$ & $18.0$ & $0.1$ & $-6$ & $7500$ & $0.0002$ & $2$ \\
$3$ & {\MgII} & $12$ & $12.02$ & $6.0$ & $17.5$ & $17.3$ & $0.1$ & $-6$ & $7900$ & $0.00004$ & $2$ \\
$4$ & {\MgII} & $50$ & $11.50$ & $3.7$ & $16.9$ & $16.8$ & $0.1$ & $-6$ & $8100$ & $0.00001$ & $2$ \\
$5$ & {\MgII} & $78$ & $11.81$ & $7.0$ & $17.3$ & $17.1$ & $0.1$ & $-6$ & $7900$ & $0.00003$ & $2$ \\
$6$ & {\MgII} & $90$ & $11.92$ & $4.7$ & $17.4$ & $17.2$ & $0.1$ & $-6$ & $7900$ & $0.00003$ & $2$ \\
$7$ & {\MgII} & $100$ & $11.77$ & $10.2$ & $17.2$ & $17.0$ & $0.1$ & $-6$ & $8000$ & $0.00002$ & $2$ \\
$8$ & {\MgII} & $140$ & $11.75$ & $9.0$ & $17.2$ & $17.0$ & $0.1$ & $-6$ & $8000$ & $0.00002$ & $2$ \\
$1$ & {\CIV} & $-8$ & $13.45$ & $19.3$ & $18.9$ & $16.0$ & $0.1$ & $-2.5$ & $18620$ & $4$ & $0.0007$ \\
$2$ & {\CIV} & $79$ & $13.69$ & $30.7$ & $19.2$ & $16.3$ & $0.1$ & $-2.5$ & $18578$ & $7$ & $0.0007$ \\
$3$ & {\CIV} & $153$ & $13.38$ & $15.8$ & $18.9$ & $16.0$ & $0.1$ & $-2.5$ & $18620$ & $3$ & $0.0007$ \\
$4$ & {\CIV} & $212$ & $14.08$ & $13.8$ & $19.3$ & $15.8$ & $0.1$ & $-2$ & $23550$ & $28$ & $0.0002$ \\
\hline
\enddata
\vglue -0.05in
\tablecomments{
\baselineskip=0.7\baselineskip
The same as Table~\ref{tab:0.8954}, except for the system at $z=0.6600$.
}
\label{tab:0.6600}
\end{deluxetable}
\clearpage

\end{document}